\documentclass[journal]{IEEEtran}

\usepackage{amsmath,amssymb,amsfonts}
\usepackage{graphicx}
\usepackage{booktabs}
\usepackage{xcolor}
\usepackage[colorlinks=true,linkcolor=black,citecolor=black,urlcolor=blue]{hyperref}
\usepackage{cite}

\graphicspath{{figures/}}
\newcommand{\papertitle}{A Probabilistic-Cost Algorithm for Large-Scale 2D Phase Unwrapping}
\newcommand{\paperauthors}{Scott Staniewicz, Geoffrey Gunter, Sara Mirzaee, Marin Govorcin, Talib Oliver-Cabrera, and Heresh Fattahi}
\newcommand{\paperkeywords}{Synthetic aperture radar, interferometry, phase unwrapping, minimum cost flow, Sentinel-1, NISAR, open-source software}

\hypersetup{
  pdftitle={\papertitle},
  pdfauthor={\paperauthors},
  pdfsubject={Two-dimensional InSAR phase unwrapping},
pdfkeywords={\paperkeywords}}

\begin{document}

\title{\papertitle}

\author{Scott Staniewicz,
  Geoffrey Gunter,
  Sara Mirzaee,
  Marin Govorcin,
  Talib Oliver-Cabrera,
  and Heresh Fattahi%
  \thanks{S. Staniewicz is with Capella Space
  (e-mail: scott.staniewicz@capellaspace.com).}%
  \thanks{S. Staniewicz, G. Gunter, H. Fattahi, S. Mirzaee, M. Govorcin, and T. Oliver-Cabrera are with the Jet Propulsion Laboratory, California Institute of Technology,
  Pasadena, CA, USA.}%
  \thanks{Manuscript submitted \today.}%
}

\markboth{
}%
{Staniewicz \MakeLowercase{\textit{et al.}}: A Probabilistic-Cost Algorithm for Large-Scale 2D Phase Unwrapping
}

\maketitle

\newcommand{\narrowfigwidth}{\columnwidth}

\begin{abstract}
  Two-dimensional phase unwrapping can dominate computation and memory usage in wide-area Interferometric Synthetic Aperture Radar (InSAR) processing.
  Here we present \emph{Whirlwind}, a minimum-cost-flow (MCF) unwrapper tailored for large interferograms of surface deformation.
  We derive fixed likelihood-ratio costs using multilook interferometric phase statistics, marginalizing uncertainty in the local phase-gradient and coherence estimates.
  A compact integer lookup table and parallel-augmenting successive-shortest-path solver with Dial bucket queues make the MCF solve efficient while avoiding flow-dependent cost relinearization.
  We test whether these improvements in speed and memory can be obtained without a loss of solution quality using multiple evaluation criteria for individual interferograms, at mission-production scale, and in end-to-end displacement time series solutions.
  On a 74-Mpixel Sentinel-1 interferogram of the 2019 Ridgecrest earthquakes, Whirlwind runs in 209~s, 28$\times$ faster than single-tile SNAPHU, while using 40\% of the peak memory.
  Across 6,014 provisional NISAR geocoded unwrapped (GUNW) products, 93\% agreed with the SNAPHU-derived results on more than 99\% of valid pixels; on seven challenging 77-MHz Antarctic frames, Whirlwind ran a median 21$\times$ faster than SNAPHU while matching 99.4\% of its integer-cycle assignments.
  Finally, in a complete Sentinel-1 time series experiment, Whirlwind improved the network-inversion consistency over SNAPHU, and in a high-resolution Capella time series of the Portuguese Bend landslide, it achieved the highest correlation with a dense GPS network among five unwrapping configurations.
  Whirlwind is provided as an open-source library capable of handling the growing volume of globally available InSAR data.
\end{abstract}


\section{Introduction}\label{sec:intro}

\IEEEPARstart{S}{atellite} InSAR has moved from an opportunistic science measurement to routine wide-area surface-deformation mapping.
Sentinel-1 and NISAR provide systematic global observations \cite{Torres2012GMESSentinel1Mission,nisar}, while services such as the European Ground Motion Service (EGMS), the Observational Products for End-Users from Remote Sensing Analysis (OPERA), Looking Into Continents from Space with Synthetic Aperture Radar (LiCSAR), and the ForM@Ter Large-Scale Multi-Temporal Sentinel-1 Interferometry Service (FLATSIM) produce regional- to continental-scale displacement products \cite{operadisp,Lazecky2020LiCSARAutomaticInSAR,Thollard2021FLATSIMForMTerLArgeScale}. The recently contracted Sentinel-1 Next Generation will increase geometric resolution and coverage \cite{Geudtner2021CopernicusSentinel1Next}.
While these observations provide tremendous opportunities for scientists and application end users, their data volumes now strain processing algorithms that were designed decades ago for individual scenes.

One of the most persistent bottlenecks in these workflows is two-dimensional phase unwrapping, which reconstructs a full, spatially consistent phase field from the interferogram's measurement of phase modulo $2\pi$ \cite{Itoh1982AnalysisPhaseUnwrapping,Ghiglia1998TwoDimensionalPhaseUnwrapping,Werner2002ProcessingStrategiesPhase}.
With many unwrapping algorithms, a single wrong correction can shift entire regions of the image by multiple cycles, leading to nontrivial errors in the final displacement.
Additionally, the problem is ill-posed, as many integer-cycle assignments are consistent with the wrapped data, and the most general problem of minimizing the number of unwrapped differences is NP-hard \cite{Chen2000NetworkApproaches}.
Phase unwrapping algorithms therefore often combine statistical estimates of noise, using inputs such as the interferometric coherence, with global topological constraints.

Approaches to two-dimensional phase unwrapping include residue-based branch cut methods \cite{Goldstein1988SatelliteRadarInterferometry} and their extensions, such as the Integrated Correlation and Unwrapping (ICU) algorithm, which uses phase gradients, amplitude, and coherence to guide cut placement before integrating phase outward from multiple seeds
\cite{buckley2000roi_pac,Hensley2002ImprovedProcessingAIRSAR}.
An alternative formulation posed by \cite{Costantini1998NovelPhaseUnwrapping} cast the integer ambiguity correction choice as a minimum-cost flow (MCF) network problem.
The Statistical-cost, Network-flow Algorithm for Phase Unwrapping (SNAPHU) added scene-dependent statistical costs and iterative network optimization
\cite{Chen2001TwoDimensionalPhaseUnwrapping,Chen2002PhaseUnwrappingLarge}, and is widely used as the default unwrapper in many open-source InSAR processing libraries.
More recently, a parallel-augmenting successive-shortest-path solver was used as the phase unwrapper for the Surface Water and Ocean Topography (SWOT) mission
\cite{Wu2017ParallelASSP,swot_pixc_atbd_2023}; an open-source implementation of this unwrapper (named ``PHASS'') is contained in the InSAR Scientific Computing Environment Version 3 (ISCE3)
\cite{isce3}\footnote{In this paper, ``ISCE3 PHASS'' refers only to this public implementation of PHASS, not to the later operational processor deployed by SWOT, which is not publicly available.}.
Further formulations unwrap on irregular networks of reliable pixels or extend the optimization in time as well as space \cite{Pepe2006ExtensionMinimumCostFlow}; see \cite{Yu2019PhaseUnwrappingReview}.
These methods make different tradeoffs between statistical expressiveness, runtime, memory, and the treatment of poorly connected regions.
In this work, we focus on regular-grid 2D phase unwrappers, which are the candidate unwrappers for large-scale missions such as the NASA ISRO SAR Mission (NISAR) and have additionally shown strong performance for OPERA Surface Displacement time series results \cite{Staniewicz2026NearRealTimeInSARPhase}.

Here, we present \emph{Whirlwind}, an open-source implementation of probabilistic-cost, 2D MCF unwrapping \cite{Costantini1998NovelPhaseUnwrapping,Carballo2000ProbabilisticCostFunctions}.  Our key contributions are (1) a fixed, directional arc cost, computed as a likelihood ratio from multilook phase statistics \cite{Lee1994IntensityPhaseStatistics} after marginalizing the estimation error in both the local phase gradient \cite{Carballo2000ProbabilisticCostFunctions} and the sample coherence \cite{Touzi1999CoherenceEstimation}; and (2) an all-integer successive-shortest-path solver \cite{Ahuja1993NetworkFlows} that combines parallel augmentation \cite{Wu2017ParallelASSP} with Dial bucket queues. Because we pose the optimization as a linear MCF instance, we can solve using a combinatorial algorithm that has substantially faster runtime than the iterative network optimization used by SNAPHU.

We evaluate whether the computational advantage of this tradeoff comes at the expense of solution quality in three ways. First, we benchmark Sentinel-1 and NISAR interferograms to compare runtime, memory usage, output coverage, and integer ambiguity solutions for five unwrapping algorithms and configurations.
Second, a campaign of 6,014 provisional NISAR products compares Whirlwind with the SNAPHU production results at a global scale, including frames with difficult displacement gradients along the coast of Antarctica.
Last, we process two end-to-end displacement time series results to compare unwrapping quality using inversion-residual consistency checks and independent GPS station velocities.

\section{Methods}\label{sec:method}

\subsection{Phase Unwrapping as Minimum-Cost Flow}\label{sec:mcf}

Let $\psi \in (-\pi,\pi]^{m\times n}$ be the wrapped phase of a multilooked interferogram. Unwrapping estimates the congruent field $\phi=\psi+2\pi K$, with $K\in\mathbb{Z}^{m\times n}$. For an oriented edge $e=(u,v)$ from pixel $u$ to adjacent pixel $v$, let the wrapped phase gradient $d_e=\operatorname{wrap}(\psi_v-\psi_u)$, where
$\operatorname{wrap}(\cdot)$ is the wrapping operator which places a value in the range $[-\pi, \pi]$.
The sum $r_p$ around a $2\times2$ pixel loop $p$ is the loop's residue \cite{Goldstein1988SatelliteRadarInterferometry}
\begin{equation}
  r_p = \frac{1}{2\pi}\sum_{e\in\partial p}s_{pe}d_e,
  \qquad r_p\in\{-1,0,+1\},
  \label{eq:residue}
\end{equation}
where $s_{pe}$ gives the orientation of edge $e$ around the loop, and the sign of a nonzero $r_p$ is its charge.
If every $r_p$ is zero, then all gradients are curl-free and we can integrate the wrapped gradients along any path to obtain an unwrapped phase field.
Nonzero residues identify topological defects, indicating that a set of $\pm2\pi$ corrections must be placed before integration.

The MCF construction uses the dual graph to find $2\pi k$ corrections for the wrapped gradients \cite{Costantini1998NovelPhaseUnwrapping}.
Each $2\times2$ loop is a node, and a graph arc crosses one pixel edge.
Positive residues supply integer flow and negative residues demand it.
The two directions across a pixel edge are kept as separate arcs, since a $+1$ and a $-1$ ambiguity difference need not carry the same statistical cost (Section~\ref{sec:cost}).

Let $A$ be the set of all directed arcs, $x_a$ be the nonnegative integer flow on arc $a$, and $c_a$ be the cost per unit flow (derived in Section \ref{sec:cost}).
Writing $\delta^+(p)$ and $\delta^-(p)$ for the arcs leaving and entering node $p$, Whirlwind solves the MCF optimization problem
\begin{equation}
  \begin{aligned}
    & \underset{ x }{\text{minimize}} & & \sum_{a\in A} c_a x_a,\\
    & \text{subject to} & & \sum_{a \in \delta^+(p)}x_a-\sum_{a \in \delta^-(p)}x_a=r_p \quad \forall p,\\
    & & &  0\le x_a\le u_a \quad \forall a \in A,
  \end{aligned}
  \label{eq:mcf}
\end{equation}
where $u_a$ is the capacity of the arc.
We model arcs whose pixel edge lies within the image as having unit capacity, $u_a=1$.
We add an outermost ring of arcs, corresponding to pixel edges just outside the image, with zero cost and unlimited capacity, $u_a = \infty$.
We also add one-sided residues along the image boundary to make the augmented graph balanced, $\sum_p r_p=0$.
The net flow across a pixel edge is its ambiguity difference $\Delta k$.
Each pixel edge is crossed by two oppositely signed arcs, $a$ and $\bar{a}$, each carrying flow $0$ or $1$; thus, $\Delta k = x^{a} - x^{\bar{a}} \in \{-1, 0, +1\}$, meaning the net ambiguity change across a pixel edge is at most one cycle.
Adding $2\pi\Delta k$ to $d_e$ cancels every residue, and we integrate to obtain $\phi$.

\subsection{Probabilistic Arc Costs}\label{sec:cost}

\begin{figure*}[!hbt]
  \centering
  \includegraphics[width=\textwidth]{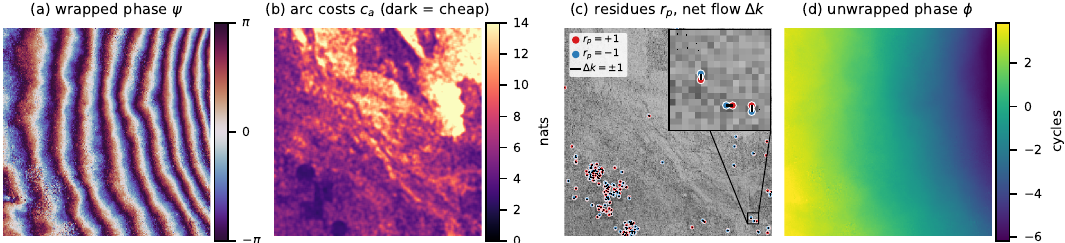}
  \caption{
    Minimum-cost-flow construction on a Sentinel-1 interferogram.
    (a) Input wrapped phase.
    (b) Arc costs evaluated from the compiled likelihood tables of Section~\ref{sec:cost} at $L{=}18$. Here we plot the minimum cost for the four arc cost arrays (up, down, left, right). Correcting coherent, low-gradient edges is expensive, while adding ambiguities to the decorrelated patch is cheap.
    (c) The residues computed using \eqref{eq:residue}. The patch contains 362 residues, leading to the 229 pixel edges with nonzero flow in \eqref{eq:mcf} (background: multilooked coherence). Note that residues concentrate where coherence is low, and each is paired with an opposite charge over a short route.
    (d) The output unwrapped phase field.
  }
  \label{fig:mechanism}
\end{figure*}

The signal slope $\alpha$ is the noise-free gradient between adjacent interferogram pixels (in radians per sample). We estimate it by applying a box filter to the wrapped phase differences, giving $\hat{\alpha} \in [-\pi, \pi]$ (note that the true slope $\alpha$ need not lie in this interval).
At each pixel, the maximum-likelihood sample coherence estimate is
\begin{equation}
  \hat\gamma \;=\;
  \frac{\left|\sum_{i=1}^{L} z_{1,i} z_{2,i}^{*}\right|}
       {\sqrt{\sum_{i=1}^{L}|z_{1,i}|^{2}}\;\sqrt{\sum_{i=1}^{L}|z_{2,i}|^{2}}},
  \label{eq:samplecoh}
\end{equation}
for the two single-look complex images $z_1, z_2$ over the $L$ looks in the multilook window; we assign the edge coherence as the minimum of its two endpoint sample coherences. Together with the effective number of looks $L$, these are the three inputs to the arc-cost model.

Following the formulation in \cite{Carballo2000ProbabilisticCostFunctions}, we assign the cost of adding one $2\pi$ correction from a log-likelihood ratio:
\begin{equation}
  c(\hat\alpha,\hat\gamma,L) \;=\;
  \max\!\left\{0, -\ln \frac{P_+}{P_0}\right\} ,
  \label{eq:llr}
\end{equation}
where $P_0$ and $P_+$ are the probabilities that the noisy unwrapped edge gradient lies within $(-\pi,\pi]$ or above $+\pi$, respectively, conditioned on $(\hat\alpha,\hat\gamma,L)$.

where $P(\Delta k = n \mid \hat\alpha,\hat\gamma,L)$ is the probability of applying an $n$-cycle correction given a slope estimate, edge coherence estimate, and effective number of looks.
Note that only $\Delta k = +1$ in the numerator, since the opposite correction is $+1$ on the reverse arc using the opposite slope $-\hat\alpha$.
The two graph directions here create the same effect as the cost pair $c^+$ and $c^-$ of \cite[eq.~(24)]{Carballo2000ProbabilisticCostFunctions}.
The $\max \{0, \cdot \}$ truncates negative costs to zero.

The probabilities in \eqref{eq:llr} follow the construction of \cite{Carballo2000ProbabilisticCostFunctions}.
Let $f_\eta(\eta\mid\gamma,L)$ be the probability density of the $L$-look interferometric phase noise $\eta$ at true coherence $\gamma$ \cite[eq.~(18)]{Lee1994IntensityPhaseStatistics}.
Rather than substitute the sample coherence $\hat\gamma$ for the true value (as in \cite{Carballo2000ProbabilisticCostFunctions}), we account for uncertainty in the coherence estimate $\hat{\gamma}$ by marginalizing the true coherence over a uniform prior on the interval $[0,1]$ using the estimator density $f(\hat\gamma\mid\gamma,L)$ of \cite{Touzi1999CoherenceEstimation}:
\begin{equation}
  \tilde f(\eta\mid\hat\gamma,L) \;\propto\;
  \int_0^1 f_\eta(\eta\mid\gamma,L)\, f(\hat\gamma\mid\gamma,L)\,d\gamma .
  \label{eq:cohmarg}
\end{equation}
This helps prevent noisy gradient estimates from being weighted too highly in the optimization.  The noise on an edge of two pixels is the difference of two phases errors, independent and symmetric, so its is the convolution $\tilde{f} \ast \tilde{f}$ with support $[-2\pi,2\pi]$ \cite[eq.~(8)]{Carballo2000ProbabilisticCostFunctions}.
For a true slope $\alpha$, the noisy unwrapped gradient is $G=\alpha+\Delta\eta$.
The conditional probabilities $q_0=P(-\pi<G\leq\pi\mid\alpha,\hat\gamma,L)$ and $q_+=P(G>\pi\mid\alpha,\hat\gamma,L)$ are
\begin{equation}
    \begin{aligned}
    q_0 &= \int_{-\pi-\alpha}^{\pi-\alpha}
      g(\Delta\eta\mid\hat\gamma,L)\,d(\Delta\eta), \\
    q_+ &= \int_{\pi-\alpha}^{\infty}
      g(\Delta\eta\mid\hat\gamma,L)\,d(\Delta\eta).
  \end{aligned}
  \label{eq:branch}
\end{equation}
These are the centered and tail integrals of \cite[eq.~(9)]{Carballo2000ProbabilisticCostFunctions}.
Since the true slope is unknown, we marginalize over $\alpha \in \mathbb{R}$ using a flat prior $p(\alpha) \propto 1$ following \cite[eq.~(14)]{Carballo2000ProbabilisticCostFunctions}.
With the Gaussian slope error approximation $f(\hat\alpha\mid\alpha,\hat\gamma)$ from \cite[eq.~(15)]{Carballo2000ProbabilisticCostFunctions}, this gives
\begin{equation}
  P_j = \int_{-\infty}^{\infty} q_j(\alpha,\hat\gamma,L)\,
  f(\hat\alpha\mid\alpha,\hat\gamma)\,d\alpha ,
  \label{eq:costmarg}
\end{equation}
for $j\in\{0,+\}$.
The Gaussian standard deviation $\sigma_\alpha$ approaches $\pi/\sqrt{3}$ as coherence falls to zero.
For numerical integration, we use $\hat\alpha\pm 8\sigma_\alpha$ and normalize the Gaussian weights over this interval.
Note that the $P_+$, as implemented, includes the entire positive tail $G>\pi$, including $G>3\pi$, where more than one correction cycle would be needed. Thus, it approximates the probability of a $+1$ correction when the mass beyond $3\pi$ is negligible.

The resulting costs are directional: the two signs of correction on the same pixel edge are not equally likely, so they are not equally priced.
Where coherence is high and the local gradient is estimated near $-\pi$, probability is concentrated near the negative wrapping boundary, so a negative correction is cheaper than a positive one; the reverse holds near $+\pi$.
A gradient estimate near zero is expensive to correct in either direction, while low-coherence edges are less trustworthy, allowing the solver to route flow through them at low cost.

Figure~\ref{fig:mechanism} shows the MCF problem construction and arc costs on a 300$\times$300-pixel portion of a Sentinel-1 interferogram.
The wrapped phase has overall high coherence, leading to high arc costs (Figure~\ref{fig:mechanism}b).
There are patches of decorrelated regions in the lower left, leading to nonzero residues (Figure~\ref{fig:mechanism}c). By balancing out the supply and demand nodes in the MCF graph, we can integrate the wrapped gradients and recover the unwrapped phase (Figure~\ref{fig:mechanism}d).

For efficient implementation, since Equation \eqref{eq:llr} depends on only three scalars, Whirlwind precomputes $P_0$ and $P_1$ on a ($\hat\alpha\times\hat\gamma\times L$) grid and interpolates at runtime. The costs are quantized by computing  $\lfloor100c_a\rfloor$ and storing them as 16-bit integers. Since these costs do not depend on flow and each interior arc carries at most one unit of flow, each arc is weighted by a single cost coefficient $c_a$ and the objective of \eqref{eq:mcf} is linear in $x$.

To aid interferograms with fast motion where a smoothed local-slope estimate can give misleading confidence near an aliased gradient (such as cryosphere interferograms with glacier motion), we add a bound on the cost model for edges above a gradient steepness threshold.
Edges whose unsmoothed phase gradient exceeds 1 radian are assigned zero cost, up to a max of 3\% of valid edges.
This empirical correction can be viewed as an approximation to a heavy-tailed slope distribution, which is simple to apply without broadly changing our likelihood derivation.
We limit the number of pixels to a few percent of the total image pixels to ensure that this empirical correction does not apply broadly to a very noisy scene.

Interferogram samples may be marked invalid by an input mask, covering water bodies, layover and shadow, or transmit gaps between subswaths in the case of NISAR's fixed pulse repetition frequency (PRF) mode.
Arcs whose pixel edge has a masked endpoint are given zero cost, so the solver can add flow across masked regions without penalty. After integration, masked pixels are set to a nodata value.
A mask that fully separates two valid regions therefore leaves their relative $2\pi$ levels undetermined, which we resolve in the bridging step of Section~\ref{sec:conncomp}.

\subsection{Integer Successive-Shortest-Path Solver}\label{sec:solver}

Since our arc costs are fixed and do not depend on flow, we can solve the MCF problem using successive shortest paths \cite[\S 9.7]{Ahuja1993NetworkFlows}.
As the overall runtime is dominated by shortest-path searches, we take advantage of the nonnegative integer costs and use Dial's bucket queues  \cite{Dial1969Algorithm360} to replace the priority queue in Dijkstra's algorithm with a circular array of first-in-first-out buckets.
The buckets are indexed by distance, and extracting the minimum distance node is performed in amortized constant time. The solver maintains node potentials, $\pi$, that keep the reduced cost of every residual arc nonnegative \cite[\S 9.3]{Ahuja1993NetworkFlows}. This permits a nonnegative-cost shortest-path search on a residual graph structure whose arcs may otherwise have negative cost.

Each iteration performs a multi-source Dijkstra search, growing a single shortest-path forest from all remaining positive residues simultaneously.
We augment one unit of flow along the forest path of the nearest negative residue in each source tree. Potentials are updated from the search distance, $\pi\gets\pi-d$.
This solver strategy is similar to the parallel-augmentation strategy described for PHASS \cite{Wu2017ParallelASSP}, related to the primal--dual method \cite[\S 9.8]{Ahuja1993NetworkFlows} without solving a maximum-flow subproblem at each iteration.

Since large masked regions can make millions of nodes equidistant, for equal-cost comparisons in the bucket queue, we break ties using first-in-first-out (breadth-first) order to favor nearby nodes and shorten correction paths.
By default, the solver performs eight full multi-source passes, then balances remaining residues with individual single-source shortest-path searches that stop at the first deficit.
If imbalance remains, it resumes multi-source augmentation before retrying the cleanup.
For any final unbalanced source and sink pairs, a final residual-graph balancing stage ignores cost to enforce feasibility.
Lastly, we integrate the wrapped gradients in integer cycle counts to avoid roundoff error and ensure the output is congruent to the input modulo $2\pi$.

\subsection{Phase preprocessing, Connected Components, and Bridging}\label{sec:conncomp}

Whirlwind also provides two optional phase preprocessing steps  which can speed up, or occasionally improve, results in interferograms with strong decorrelation. Similar to \cite{Chen2015PersistentScattererInterpolation,Wang2022AccuratePersistentScatterer}, we include a spiral interpolator which replaces low-weight pixels with a distance-weighted average of nearby high-weight unit phasors.
Alternatively, we include an adaptive power-spectrum filter (a ``Goldstein filter'' \cite{Goldstein1998RadarInterferogramFiltering}) which smooths the wrapped interferogram in overlapping Fourier-domain patches. When using either preprocessing step, the modified phase is used only to estimate the integer ambiguity field $K$; after unwrapping, the integer solutions are transferred to the original wrapped phase. The returned samples therefore retain the original phase and mask.

In addition to the unwrapped phase field, Whirlwind also outputs a grid of connected-component labels (self-contained regions of reliably unwrapped phase) using an algorithm similar to SNAPHU's \cite{Chen2002PhaseUnwrappingLarge}. After unwrapping, a pixel edge is reliable if perturbing its ambiguity by $\pm 1$ cycle would raise the MCF objective by more than a preset threshold (since high-coherence edges have larger correction cost). Labels are grown over reliable edges using a breadth-first search.

When the valid mask splits the scene into disconnected regions (e.g., rivers, coastlines, missing tiles), the wrapped phase carries no information about their relative $2\pi$ levels, and the solver unwraps each region independently. Therefore, by default we include a post-processing step to ``bridge'' integer shifts between regions. Our algorithm is similar to those in ISCE3 and MintPy \cite{isce3,Yunjun2019MintPy}, but constructs the spanning tree in order of decreasing region size avoids noisier small islands from disrupting the solutions of larger land masses. Each successively smaller region is attached to its nearest already visited region using the closest pair of boundary pixels. For each bridging tree edge, we take the median unwrapped phase of 65 pixel windows around the two endpoints, round their difference to integer cycles, and propagate the offset from parent to child.

\section{Results}\label{sec:results}
\begin{figure*}[!t]
  \centering
  \includegraphics[width=\textwidth]{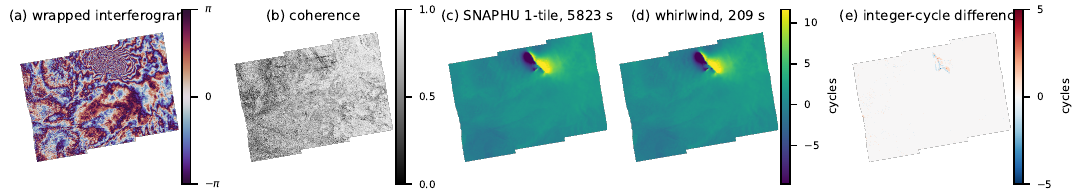}
  \caption{
    Coseismic interferogram from July 4, 2019 to July 10, 2019, spanning both the M6.4 and M7.1 Ridgecrest earthquakes showing
    (a) wrapped interferogram phase,
    (b) sample coherence,
    (c) SNAPHU single-tile unwrapped phase,
    (d) Whirlwind unwrapped phase, (e) integer-cycle difference (d)$-$(c)
  }
  \label{fig:ridgecrest}
\end{figure*}

We evaluate Whirlwind in three ways:
Sections~\ref{sec:ridgecrest} and \ref{sec:nisar-beta} measure runtime, memory, and integer cycle agreement against SNAPHU and other common unwrappers on individual interferograms.
Section~\ref{sec:campaign} extends this to a global comparison of 6,014 provisional NISAR GUNW products, measuring agreement with the production SNAPHU settings.
Sections~\ref{sec:miami} and \ref{sec:palosverdes} then compare full time series produced with multiple unwrappers using consistency checks on the time series inversion residuals and accuracy compared to GPS velocities.

\subsection{The July 2019 Ridgecrest Earthquakes}\label{sec:ridgecrest}

We first compare the unwrapping results of Whirlwind to SNAPHU on a full-frame Sentinel-1 coseismic interferogram of the M6.4 and M7.1 Ridgecrest earthquakes \cite{Ross2019Ridgecrest,Xu2020CoseismicDisplacementsSurface}. We used geocoded coregistered single-look complex (CSLC) products created by the OPERA project \cite{operadisp} from descending track T064 between 2019-07-04 and 2019-07-10 (using the 27 Burst IDs contained in OPERA DISP-S1 Frame 16941 \cite{Staniewicz2026NearRealTimeInSARPhase}).  We cross-multiplied the CSLCs, multilooked $3 \times 6$ in the $Y$- and $X$-directions, and stitched the burst interferograms on the products' UTM grid to form a $7825\times 9509$-pixel interferogram at 30-m posting (Figure~\ref{fig:ridgecrest}a-b). We unwrapped the phase using the sample coherence with Whirlwind, single-tile SNAPHU, and tiled SNAPHU \cite{Chen2002PhaseUnwrappingLarge} configured as $5 \times 5$ tiles with four tiles running in parallel.

Whirlwind unwraps the frame in a single tile in 209~s (3.5 minutes) and uses 8.7~GB of RAM. Single-tile SNAPHU takes 5823~s (97 minutes) and 21.2~GB; the $5\times 5$ tiled configuration with single-tile reoptimization takes 1166~s and 10.2~GB.
Tiled SNAPHU reproduces its own single-tile output exactly due to the single-tile reoptimization step included in SNAPHU.
Whirlwind agrees with the single-tile SNAPHU solution over 99.7\% of pixels after integer-cycle alignment; the 0.3\% of differences occur near the surface rupture where the wrapped gradient is steepest (Figure~\ref{fig:ridgecrest}c--e).
Further analysis of the residuals near the rupture are shown in Section~\ref{sec:discussion}.

\begin{figure*}[!t]
  \centering
  \includegraphics[width=\textwidth]{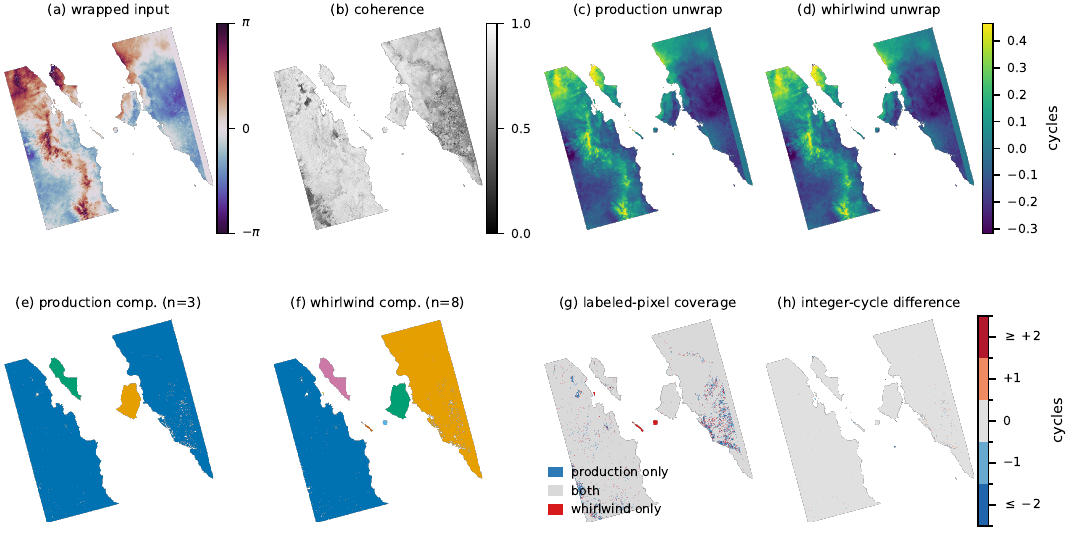}
  \caption{
    Input components to global NISAR comparison on one example frame on the coastal Gulf of California frame C3\_T5\_F16 (2025-10-17/2025-10-29):
    (a) re-wrapped production phase, the input handed to each unwrapper;
    (b) coherence;
    (c) production unwrapped phase;
    (d) Whirlwind unwrapped phase after a single global $2\pi$ alignment;
    (e)-(f) connected-component labels from each unwrapper (colors arbitrary, ordered by component size);
    (g) pixels that only one unwrapper assigns a component label;
    (h) per-pixel integer-cycle difference
  }
  \label{fig:nisar-overview}
\end{figure*}

\begin{figure}[!t]
  \centering
  \includegraphics[width=\narrowfigwidth]{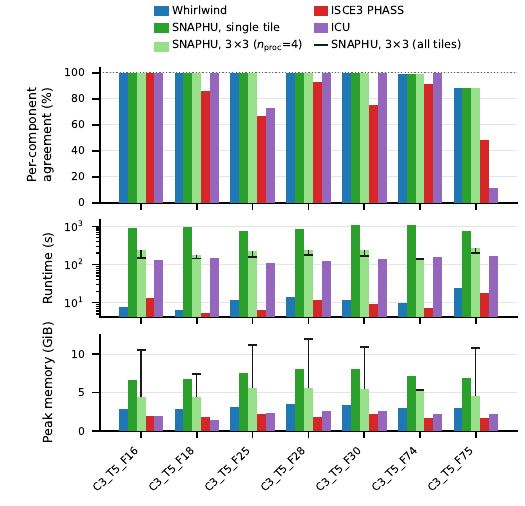}
  \caption{
    NISAR Beta GUNW unwrapper comparisons for Whirlwind (blue), single-tile SNAPHU (dark green), $3{\times}3$-tiled SNAPHU with reoptimization (light green), ISCE3 PHASS (red), and ICU (purple).
    (top) Per-component integer-cycle agreement with the SNAPHU production unwrapping results,
    (middle) runtime (in seconds) of the unwrapping configurations,
    (bottom) peak RAM used by unwrapper.
    Tiled SNAPHU memory and runtimes using maximum parallelism (improving runtime, but increasing memory usage) are shown using stems.
  }
  \label{fig:results-beta-summary}
\end{figure}

\begin{table}[!t]
  \centering
  \caption{Runtime and peak memory over the 13 frames ($\sim$18--21~Mpixel each)
  }
  \label{tab:perf}
  \begin{tabular}{lrr}
    \toprule
    Unwrapper & Runtime (s) & Peak memory (GB) \\
    \midrule
    Whirlwind                & 6--24      & 2.7--3.5 \\
    SNAPHU, single tile      & 466--1242  & 6.2--8.1 \\
    SNAPHU, $3{\times}3$ ($n_{\rm proc}{=}4$) & 100--268 & 4.4--5.6 \\
    SNAPHU, $3{\times}3$ (all tiles) & 97--201 & 5.3--12.0 \\
    ISCE3 PHASS              & 5.5--23    & 1.6--2.2 \\
    ICU                      & 109--204   & 1.4--2.6 \\
    \bottomrule
    \multicolumn{3}{@{}p{\columnwidth}@{}}{}
  \end{tabular}
\end{table}

\subsection{NISAR Beta Benchmark}\label{sec:nisar-beta}

Thirteen NISAR Beta GUNW frames \cite{nisar_l2_gunw_beta_v1} are benchmarked by re-wrapping the production HH, 80-meter unwrapped phase, then unwrapping with multiple algorithms and configurations. We identify frames by mission cycle, track, and frame number from the NISAR product name (\texttt{NISAR\_L2\_PR\_GUNW\_003\_005\_A\_016\_\ldots}). All 13 Beta frames span 2025-10-17--2025-10-29 and are from cycle~3 (e.g., C3\_T5\_F16 is cycle~3, track~5, frame~16).

Figure~\ref{fig:nisar-overview} shows the inputs and outputs used in the comparison for each frame. The inputs are re-wrapped phase and the coherence layer provided in the GUNW (Figure~\ref{fig:nisar-overview}a-b), as well as the subswath mask to avoid invalid pixels.
We compare the production unwrapped phase (Figure~\ref{fig:nisar-overview}c) to the Whirlwind unwrapped phase (Figure~\ref{fig:nisar-overview}d) after adjusting the per-connected-component areas (regions of spatially continuous, internally consistent unwrapped pixels as identified by SNAPHU) to a median alignment within each component.
Note that here we are comparing to the SNAPHU baseline rather than an absolute accuracy metric.
We also compare the connected component labels (Figure~\ref{fig:nisar-overview}e-f), as well as the total coverage (defined as the number of pixels labelled with a valid, nonzero connected component label).
Agreement is computed as integer-cycle match, both with one global offset added, as well as with a per-connected-component alignment to avoid cases where bridging algorithms disagree (Figure~\ref{fig:nisar-overview}h).
For 13 Beta NISAR frames, we unwrapped all frames with both whirlwind and four other unwrapper configurations: SNAPHU in single-tile mode, SNAPHU using a $3 \times 3$ tiling scheme, the PHASS phase unwrapper contained in ISCE3, and the ICU unwrapper included in ISCE2.

We show a summary of the Beta frame comparisons in (Figure~\ref{fig:results-beta-summary}) for the 7 of the frames out of the 13 tested with the largest unwrapper differences.
Whirlwind shows strong agreement with production at 98.8--100\% on 12 of 13 frames (Figure~\ref{fig:results-beta-summary}, blue bars).
ISCE3 PHASS and ICU both drop below 90\% agreement on four of the frames due to differences in the algorithmic handling of low correlation or fragmented island scenes.
Note that on the 13th frame, our own configuration of SNAPHU only reaches an agreement of 88\% with production, possibly due to the differences between unwrapping the radar-coordinate interferogram and geocoding after with the re-wrapped geocoded interferogram.
For the other 6 frames not included in Figure~\ref{fig:results-beta-summary}, all unwrappers agree at 90\% or better.

Both Whirlwind and PHASS show the strongest overall performance in terms of runtime and peak memory (Table~\ref{tab:perf}). Whirlwind runs the whole-frame solve $\sim$30-200$\times$ faster than the single-tiled SNAPHU configuration (median 92$\times$) and 6-21$\times$ faster than tiled SNAPHU at full tile concurrency, using under half the memory of single-tile SNAPHU. Adding more concurrency to the tiled SNAPHU mode offers some speedup, but increases memory usage (1-6 GB).
We attribute part of the disagreement between SNAPHU and PHASS to PHASS's target object of unwrapping SWOT interferograms.  SWOT aims to unwrap and isolate many separate bodies of water, where a splintering of the final unwrapped phase and a more aggressive masking is desirable.

\begin{figure*}[!t]
  \centering
  \includegraphics[width=\textwidth]{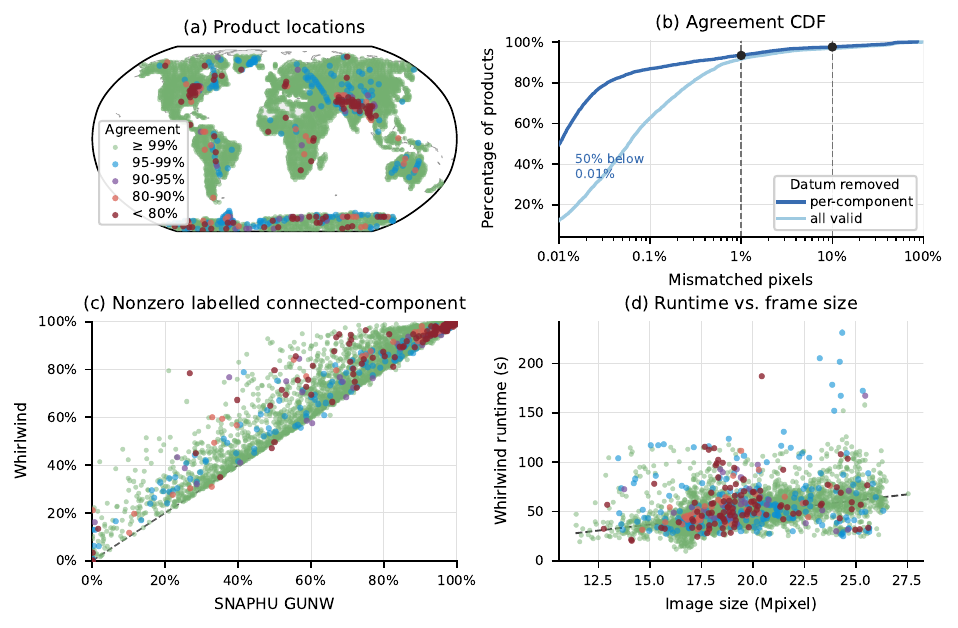}
  \caption{
    Provisional NISAR campaign.
    (a) Centers of the 6,014 evaluated products, colored by per-component agreement with production.
    (b) Cumulative distribution of the mismatch rate. Vertical lines mark 1\% and 10\% mismatch.
    (c) Fraction of valid pixels assigned a connected-component label compared to nonzero connected component from GUNW layer.
    (d) Whirlwind runtime vs. frame size. Dashed line is median rate.
  }
  \label{fig:campaign}
\end{figure*}

\subsection{Provisional NISAR Campaign}\label{sec:campaign}

We next ran only Whirlwind over 6,014 provisional GUNW products from the July 2026 release with at least 25\% land coverage (Figure~\ref{fig:campaign}a)  \cite{nisar_l2_gunw_provisional_v1}. The dataset contains 3,042 ascending and 2,972 descending acquisitions, with image sizes of 11--28~Mpixel based on the acquisition modes included and the total swath coverage.
As described in Section~\ref{sec:nisar-beta}, we re-wrapped the production unwrapped phase, used the product subswath mask and coherence, and measured the per-production-component integer-cycle match.

Median agreement is 99.99\% (0.010\% mismatched pixels), and over 93\% of frames tested agree on at least 99\% of pixels (Figure~\ref{fig:campaign}b).
Whirlwind assigns a nonzero component label to a median 95\% of valid pixels, compared with 91\% in the production settings of SNAPHU (Figure~\ref{fig:campaign}c), and the median runtime of Whirlwind is 47 seconds, with a 95th percentile runtime of 77 seconds and median peak memory of 4.34~GB (95th percentile 5.60~GB).
Note that many of the largest disagreements (agreement $<80\%$, Figure~\ref{fig:campaign}a) are not due to the underlying unwrapping algorithms, but with challenges with the data. Many low agreement frames cluster over the US Midwest and India, where NISAR acquires in its quad-polarization fixed-PRF mode, and over Antarctica, where high phase gradients and low correlation lead to challenging unwrapping conditions for any phase unwrapper. In the fixed PRF mode, transmit gaps split the scene into disconnected valid regions which must then be reconnected in post-processing. Whirlwind's preprocessing and bridging steps  (Section~\ref{sec:conncomp}) appear to handle this better than the production unwrapper configuration. Further examples of the frames with the largest disagreement between Whirlwind and SNAPHU are given in Supplement~1.

\begin{figure}[!t]
  \centering
  \includegraphics[width=\narrowfigwidth]{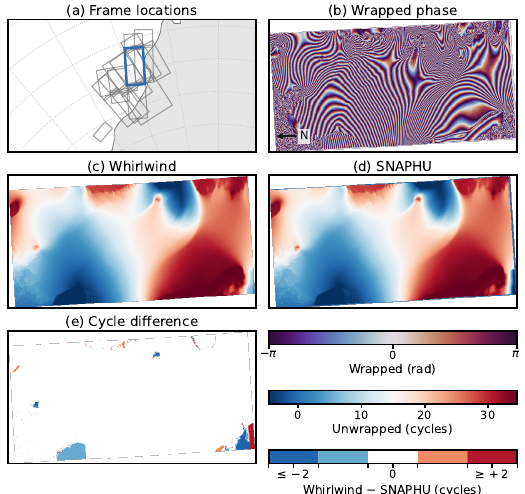}
  \caption{
    Antarctic outlet-glacier test:
    (a) Provisional NISAR GUNW footprints. Blue highlighted frame is shown in subsequent panels.
    (b) Re-wrapped production phase (rotated so north is to the left).
    (c) Whirlwind and (d) production SNAPHU unwrapped phase.
    (e) Integer-cycle difference between (c) and (d).
  }
  \label{fig:antarctica}
\end{figure}

\subsection{Antarctic Outlet Glaciers}\label{sec:antarctica}

The provisional NISAR archive \cite{nisar_l2_gunw_provisional_v1} includes 77~MHz frames over fast-flowing outlet glaciers in Antarctica. The interferograms contain a combination of dense, near-aliased fringes with decorrelated water regions (Figure~\ref{fig:antarctica}b). We benchmark seven frames of sizes 14-25 million pixels by re-wrapping the production phase and running both Whirlwind and SNAPHU on the same product mask and coherence.

Whirlwind agrees with the production unwrap over a median 99.4\% of pixels (94--100\%) and runs a median 21$\times$ faster (12--132$\times$).
The combined runtime of all seven frames for Whirlwind is 4.4~min, compared with 1.9~h for SNAPHU (Table~\ref{tab:antarctica}), while showing very high agreement (Figure~\ref{fig:antarctica}e).

\begin{table}[!b]
  \centering
  \caption{Runtime and agreement for seven 77~MHz Antarctic GUNW frames.}
  \label{tab:antarctica}
  \begin{tabular}{lrr}
    \toprule
    Unwrapper & Runtime (s) & Agreement (\%) \\
    \midrule
    Whirlwind       & 8--79 (med.\ 46)      & 94--100 (med.\ 99.4) \\
    SNAPHU (1 tile) & 552--1386 (med.\ 972) & 95--100 (med.\ 99.9) \\
    \bottomrule
  \end{tabular}
\end{table}

\begin{figure*}[!t]
  \centering
  \includegraphics[width=\textwidth]{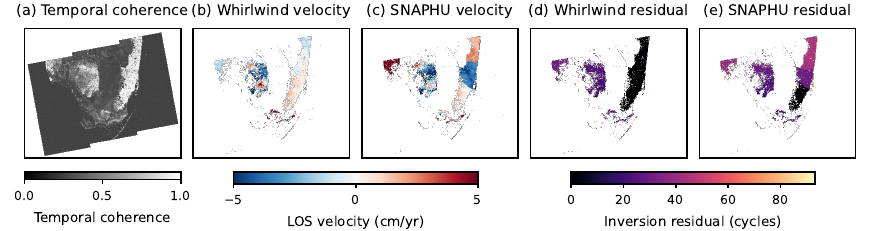}
  \caption{South Florida Sentinel-1 time series comparison.
    (a) Temporal coherence from phase linking.
    (b) Whirlwind and (c) SNAPHU average line-of-sight velocity (in cm/year) from the same set of phase-linked input interferograms. Red indicates motion toward the satellite, blue is motion away.
    (d) Whirlwind and (e) SNAPHU network-inversion residual; any nonzero value indicates a likely unwrapping error in at least one interferogram.
  Panels (b)--(e) show pixels with temporal coherence above 0.7.}
  \label{fig:miami}
\end{figure*}

\subsection{Sentinel-1 Time Series over South Florida}\label{sec:miami}

The benchmarks above compare different unwrapper outputs to each other.
As a complementary check, we can check the consistency of a redundant set of phase-linked interferograms with zero expected wrapped phase misclosure.
Because phase linking estimates a consistent wrapped phase vector (i.e. with no phase misclosure), any nonzero residual left by the $L_1$ network inversion comes from integer unwrapping errors in at least one interferogram \cite{Staniewicz2026NearRealTimeInSARPhase}.
While the converse does not hold (zero residuals do not imply zero unwrapping errors), we can compare the results of unwrapping methods on the same phase linked interferograms to see which leaves fewer residuals (and therefore has a higher likelihood of accurate phase unwrapping).

We applied this test using a stack of Sentinel-1 data over south Florida in a challenging region containing Everglades wetlands, canal networks, and barrier islands which split the scene into multiple coherent areas (Figure~\ref{fig:miami}a).
We processed 26 Sentinel-1 OPERA CSLC acquisitions spanning 2024 with the OPERA production Dolphin software \cite{Staniewicz2024DolphinPythonPackage} to perform phase linking nearest-3 interferogram formation. We created 72 interferograms at 30-m posting (approximately 6,000 $\times$ 8,000 pixels), and applied interpolation of low-quality pixels with phase similarity below 0.4 before unwrapping \cite{Chen2015PersistentScattererInterpolation}.
We unwrapped the identical interferograms with the tiled SNAPHU configuration ($5\times5$ tiles, with a single-tile reoptimization to avoid stitching artifacts) and with Whirlwind.
We inverted each unwrapped network with the same $L_1$ inversion to obtain displacement time series and average line-of-sight velocity. We performed no further filtering or atmospheric corrections.

Figure~\ref{fig:miami} shows the average velocity and the summed absolute inversion residual for both unwrappers over the 3.9 million pixels with temporal coherence above 0.7.
In the largest urban coherent area, Whirlwind's unwrapped network inverts with zero residual (Figure~\ref{fig:miami}d, right side), while SNAPHU's accumulates $\sim$30 cycles of misclosure across the network.
The breaks along the map of SNAPHU residual occur at canal boundaries, highlighting the effectiveness of the built-in bridging algorithm for Whirlwind, which shows no artifacts on the east coast of the velocity map.
We note that both unwrappers contain errors on the opposite coast of Florida across the Everglades: 43\% of the coherent pixels show at least one cycle of total misclosure for Whirlwind, versus 82\% for SNAPHU.

The runtime difference on this Sentinel-1 stack shows similar relative performance to the NISAR benchmarks. Unwrapping all interferograms with the $5 \times 5$ tiled configuration of SNAPHU took 26 hours using 10 CPUs (two interferograms and ten total tile processes running concurrently). Whirlwind took 6~hours total using 4 CPUs, whose lower memory and CPU usage also permit more interferograms to be unwrapped concurrently on the same machine.
Since unwrapping often dominates the wall time of end-to-end time series processing \cite{Staniewicz2026NearRealTimeInSARPhase}, the reduction in CPU usage and memory usage allows both lower latency products, as well as cheaper cloud-computing instances.

\begin{figure*}[!t]
  \centering
  \includegraphics[width=\textwidth]{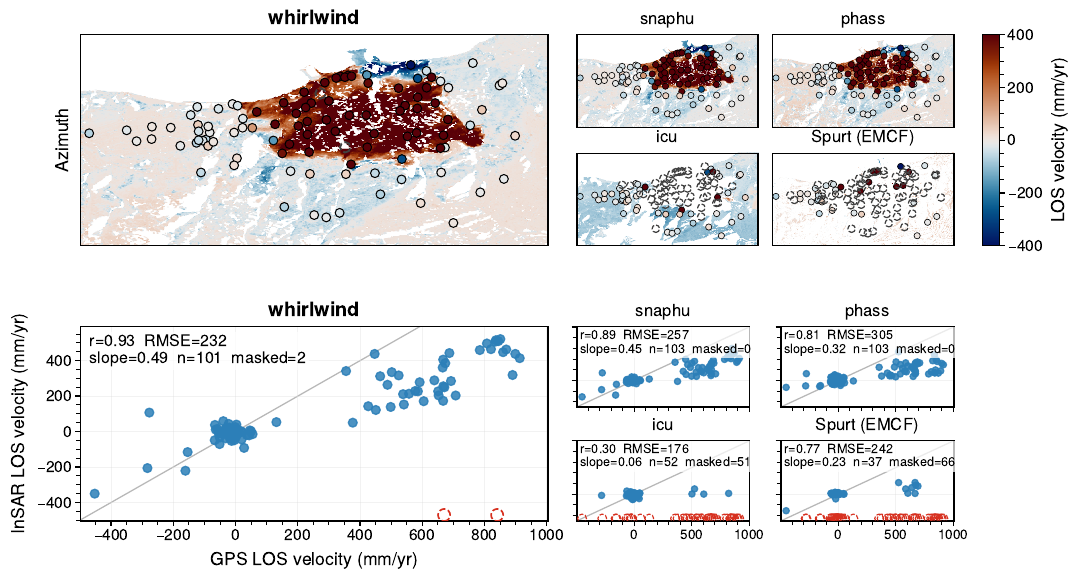}
  \caption{Portuguese Bend landslide case study:
    Whirlwind (large panels) and four other unwrapping algorithms (insets).
    \emph{top}: line-of-sight (LOS) velocity maps in radar coordinates.
    GPS overlaid and filled by their survey LOS velocity on the same scale; black-edged markers are the stations scored in the regression.
    Blank pixels were output as NaN or marked as a zero connected component label.
    \emph{Bottom}: InSAR LOS velocity versus GPS average rates, projected onto LOS. Red markers along the bottom edge mark stations where unwrapper was masked.
    Diagonal line indicates 1-1 line.
  }
  \label{fig:palosverdes}
\end{figure*}

\subsection{Portuguese Bend Landslide}\label{sec:palosverdes}

As a final test, we processed a stack of high-resolution ($\sim$ 50 centimeter) X-band data capturing fast-moving surface deformation, at or beyond the resolving limits of InSAR without resorting to amplitude-based methods (e.g., pixel offset tracking) \cite{Pepin2024AliasingInSAR}.
The Portuguese Bend landslide (Palos Verdes, California) has moved continuously for decades at rates near 1~m/yr \cite{Calabro2010PortugueseBend}.
We used 52 Capella X-band spotlight images ($\sim$3-day repeat, November 2025--May 2026), phase-linked with Dolphin \cite{Staniewicz2024DolphinPythonPackage,Ansari2017SequentialEstimator} into a nearest-3 network of 150 multilooked interferograms ($3802\times4065$ pixels).
We reran the unwrapping portion of the time series workflow with each algorithm on identical input interferograms and coherence images: Whirlwind; SNAPHU with $2 \times 2$ tiling (4 parallel tiles); ISCE3 PHASS; ICU; and a three-dimensional extended-minimum-cost-flow (EMCF) unwrapper \cite{Agram2022EfficientGlobalScale,Pepe2006ExtensionMinimumCostFlow} implemented in the Spurt library \cite{spurt}.
The 150 unwrapped pairs from each unwrapper are inverted to a 51-epoch displacement series using the L1 inversion in Dolphin.
Locations where an unwrapper outputs a masked pixel or zero connected-component label in more than 100 of the 150 interferograms are excluded from the comparison statistics.
After inversion, each velocity map is re-referenced to a common stable urban block.
The InSAR velocity at a station is the median within a $5$ meter window centered on its radar coordinates to account for GPS-to-pixel misalignment from geolocation errors.

\begin{table}[!t]
  \centering
  \caption{Portuguese Bend unwrapper comparison.
    LOS velocity vs.\ 103 GPS stations.
    $n$ = stations retained by each unwrapper.
    slope is the regression of InSAR on GPS.
  Wall time is for the total unwrapping time for 150 interferograms.}
  \label{tab:palosverdes}
  \begin{tabular}{lrrrrr}
    \toprule
    Unwrapper & $n$ & $r$ &  slope & Runtime  \\
    \midrule
    Whirlwind            & 101 & 0.93 &  0.49 & 0.6~h \\
    SNAPHU   & 103 & 0.89 & 0.45 & 8.0~h \\
    PHASS          & 103 & 0.81 & 0.32 & 1.7~h \\
    ICU                  &  52 & 0.30 &  0.06 & 0.7~h \\
    Spurt (EMCF)         &  37 & 0.77 &  0.23 & 2.0~h \\
    \bottomrule
  \end{tabular}
\end{table}

Whirlwind and SNAPHU yield GPS correlations of \(r=0.93\) and \(r=0.89\), respectively, while requiring 0.6 h and 8.0 h of total unwrapping time (Figure~\ref{fig:palosverdes}; Table~\ref{tab:palosverdes}).
The runtime of PHASS here is roughly 3$\times$ slower than Whirlwind, and PHASS masks more pixels within individual interferograms. Similar to Section~\ref{sec:nisar-beta}, PHASS masks more pixels due to algorithm's target application of unwrapping many separate water bodies.
The branch cut algorithm of ICU retains the fewest valid pixels of 2D unwrappers, covering only 52 of 103 stations.
Spurt's sparse 3D EMCF attains $r=0.77$ at 37 stations, likely indicating that there was not sufficient spatial coverage to capture the fast deformation gradient of the landslide.

\section{Discussion and Limitations}\label{sec:discussion}

\begin{figure}[!t]
  \centering
  \includegraphics[width=\columnwidth,height=0.65\textheight,keepaspectratio]{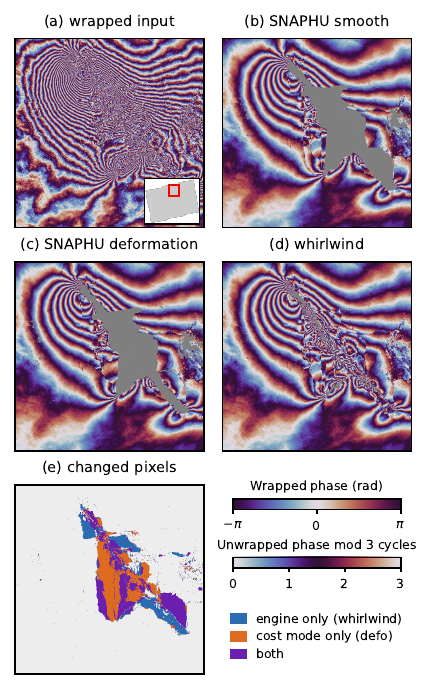}
  \caption{SNAPHU cost mode sensitivity, zooming into Ridgecrest near-fault crop (red box in the inset):
    (a) wrapped input phase;
    (b) smooth-cost SNAPHU
    (c) deformation-cost SNAPHU, and
    (d) Whirlwind. Unwrapped outputs are re-wrapped modulo three cycles to show differences more clearly.
    Locations with a zero connected component label are masked in gray.
    (e) Change map of pixel differences, marking where Whirlwind vs. SNAPHU smooth (blue), SNAPHU smooth vs. deformation (orange), or both (purple) changed the integer cycle solution.
  }
  \label{fig:deforewrap}
\end{figure}

\subsection{Runtime improvements over SNAPHU}\label{sec:runtime-discussion}

Whirlwind agrees with SNAPHU at a large majority of tested configurations, yet runs one to two orders of magnitude faster.
This comes from an intentional narrowing of the scope of the problem.
SNAPHU \cite{Chen2001TwoDimensionalPhaseUnwrapping} generalized the network formulation to maximum-a-posteriori (MAP) estimation problem, where each arc carries a cost function $g_e(x_e)$ derived from statistical radar models. There are different models used for topographic interferograms, ``deformation'' interferograms with possible discontinuities, and ``smooth'' interferograms with no expected discontinuities. Its topography and deformation modes are non-convex in $x_e$, and the general problem is NP-hard \cite{Chen2000NetworkApproaches}.
Its smooth mode, used by many surface deformation mapping processors including the NISAR production configuration, is convex; however, the problem is still nonlinear, as the costs depend on the current flow in the graph edges. SNAPHU therefore uses an iterative, nonlinear solver that augments flow and re-evaluates the incremental costs around the current solution \cite{Chen2002PhaseUnwrappingLarge}.
Whirlwind does not have a specialized topography mode or discontinuous deformation like SNAPHU. The design goal is a probabilistic cost model statistically similar to SNAPHU's for interferograms of surface deformation, but linear in the flow, so that it can take advantage of existing efficient MCF algorithms.

\subsection{Comparison to SNAPHU's Deformation mode for coseismic rupture}

All comparisons shown previously with SNAPHU have used the ``smooth'' cost mode, including the Ridgecrest earthquake from Section~\ref{sec:ridgecrest}.
The deformation-mode costs add a discontinuity parameter that limits the cost growth across large jumps in the phase gradient \cite{Chen2001TwoDimensionalPhaseUnwrapping}.

To check if Whirlwind performs noticeably worse in this regime, we compared results to SNAPHU with deformation costs (with a maximum cycle break of 1.2 cycles, the default) on the Ridgecrest frame of Section~\ref{sec:ridgecrest} (Figure~\ref{fig:deforewrap}).
Changing the cost mode produced minor solution differences close to the surface rupture (Figure~\ref{fig:deforewrap}b, c).
Both SNAPHU results assigned a zero connected-component label to the area near the rupture where coherence is low.
The deformation-mode ran $\sim1.5\times$ slower than the smooth configuration but assigned slightly fewer nonzero component labels (95.5\% vs.\ 96.5\% of valid pixels).
While some regions unwrapped by Whirlwind may be incorrect, it is ambiguous which unwrapper (and cost mode) produces the best result in this highly complex interferogram.
We note that due to this ambiguity and the inherent noise levels of the wrapped phase near the surface rupture, many studies of the earthquake source avoid unwrapping entirely. Alternatives to phase unwrapping for this application include analyzing phase gradients
\cite{Xu2020CoseismicDisplacementsSurface}, fitting geophysical models
directly to the wrapped phase \cite{Feigl2009MethodModellingRadar}, or using pixel offset tracking \cite{Strozzi2002GlacierMotionEstimation}.

\section{Conclusion}\label{sec:conclusion}

We have developed an efficient phase unwrapper, Whirlwind, designed to unwrap large-scale interferograms with low memory usage and predictable runtimes.
Whirlwind chooses a fixed cost structure for its MCF optimization, enabling fast MCF solver algorithms on large problems.
We demonstrate using Sentinel-1, Capella Space, and over 6,000 frames of NISAR data that Whirlwind may provide one to two orders of magnitude runtime improvement over single-tiled SNAPHU with comparable unwrapped phase solutions.
Whirlwind is provided as an open source library both for local use and for mission-scale production.

\section*{Acknowledgements}

Whirlwind has an open source implementation at \url{https://github.com/scottstanie/whirlwind-insar}.
Sentinel-1 OPERA CSLC and NISAR Level-2 GUNW data are available from
the Alaska Satellite Facility.
The Capella Space Portuguese Bend SLCs are available through the Capella Open Data catalog on AWS at \url{https://registry.opendata.aws/capella_opendata}.

The Portuguese Bend GPS survey data are from McGee Surveying Consulting, ``Portuguese Bend Landslide Monitoring---Movement Data Posting No.~2--M92, Sept.~17, 2025--Mar.~4, 2026,'' prepared for the City of Rancho Palos Verdes, Mar.~2026, available at \url{https://www.rpvca.gov/1426/Landslide-Surveys-and-Mapping}.
We thank Alexander Handwerger for processing and providing the Rancho Palos Verdes GPS data.

This work was largely supported by the OPERA and NISAR projects. Funded by the Satellite Needs Working Group, OPERA develops remote sensing products to address Earth observation needs across U.S. civilian federal agencies. NISAR project operates NISAR mission and produces analysis ready products from NISAR observations. The OPERA and NISAR projects are managed by the Jet Propulsion Laboratory, California Institute of Technology, under a contract with the National Aeronautics and Space Administration (80NM0018D0004) © 2026 Jet Propulsion Laboratory, California Institute of Technology. Government sponsorship acknowledged. 

\IEEEtriggeratref{32}
\bibliographystyle{IEEEtran}
\bibliography{references}

@article{Touzi1999CoherenceEstimation,
  title = {Coherence Estimation for {SAR} Imagery},
  author = {Touzi, Ridha and Lopes, Armand and Bruniquel, Jerome and
            Vachon, Paris W.},
  year = {1999},
  journal = {IEEE Transactions on Geoscience and Remote Sensing},
  volume = {37},
  number = {1},
  pages = {135--149},
  doi = {10.1109/36.739146}
}

@article{Strozzi2002GlacierMotionEstimation,
  title = {Glacier Motion Estimation Using {{SAR}} Offset-Tracking Procedures},
  author = {Strozzi, T. and Luckman, A. and Murray, T. and Wegmuller, U. and Werner, C.L.},
  year = 2002,
  month = nov,
  journal = {IEEE Transactions on Geoscience and Remote Sensing},
  volume = {40},
  number = {11},
  pages = {2384--2391},
  issn = {0196-2892},
  doi = {10.1109/TGRS.2002.805079},
  urldate = {2026-09-16},
  langid = {english}
}

@article{Chen2000NetworkApproaches,
  title = {Network Approaches to Two-Dimensional Phase Unwrapping:
           Intractability and Two New Algorithms},
  author = {Chen, Curtis W. and Zebker, Howard A.},
  year = {2000},
  journal = {Journal of the Optical Society of America A},
  volume = {17},
  number = {3},
  pages = {401--414},
  doi = {10.1364/JOSAA.17.000401}
}

@article{Chen2001TwoDimensionalPhaseUnwrapping,
  title = {Two-Dimensional Phase Unwrapping with Use of Statistical Models for
           Cost Functions in Nonlinear Optimization},
  author = {Chen, Curtis W. and Zebker, Howard A.},
  year = {2001},
  journal = {Journal of the Optical Society of America A},
  volume = {18},
  number = {2},
  pages = {338--351},
  doi = {10.1364/JOSAA.18.000338}
}

@book{Ghiglia1998TwoDimensionalPhaseUnwrapping,
  title = {Two-Dimensional Phase Unwrapping: Theory, Algorithms, and Software},
  author = {Ghiglia, Dennis C. and Pritt, Mark D.},
  year = {1998},
  publisher = {Wiley},
  address = {New York},
  isbn = {978-0-471-24935-1}
}

@techreport{swot_pixc_atbd_2023,
  title = {Surface Water and Ocean Topography Project Algorithm Theoretical Basis Document: {{Level}} 2 {{KaRIn}} High Rate Pixel Cloud Science Algorithm Software ({{L2}}\_{{HR}}\_{{PIXC}})},
  author = {{Jet Propulsion Laboratory and Centre National d'\'Etudes Spatiales}},
  year = 2023,
  month = jul,
  number = {D-105504},
  institution = {{Jet Propulsion Laboratory and Centre National d'\'Etudes Spatiales}}
}

@techreport{buckley2000roi_pac,
  title = {{{ROI}}\_{{PAC}} Documentation---Repeat Orbit Interferometry Package},
  author = {Buckley, S. M. and Rosen, P. A. and Persaud, Patricia},
  year = 2000,
  institution = {Jet Propulsion Laboratory},
  address = {Pasadena, CA}
}

@inproceedings{Hensley2002ImprovedProcessingAIRSAR,
  title = {Improved {{Processing}} of {{AIRSAR Data Based}} on the {{GeoSAR Processor}}},
  author = {Hensley, Scott and Chapin, Elaine and Freedman, Adam and Michel, Thierry},
  year = 2002,
  booktitle = {Proceedings of the 2002 AIRSAR Earth Science and Applications Workshop},
  address = {Pasadena, CA},
  langid = {english}
}

@article{Agram2022EfficientGlobalScale,
  title = {An {{Efficient Global Scale Sentinel-1 Radar Backscatter}} and {{Interferometric Processing System}}},
  author = {Agram, Piyush S. and Warren, Michael S. and Calef, Matthew T. and Arko, Scott A.},
  year = 2022,
  month = jul,
  journal = {Remote Sensing},
  volume = {14},
  number = {15},
  pages = {3524},
  doi = {10.3390/rs14153524},
  urldate = {2026-07-10},
  langid = {english}
}

@book{Ahuja1993NetworkFlows,
  title = {Network Flows: Theory, Algorithms, and Applications},
  author = {Ahuja, Ravindra K. and Magnanti, Thomas L. and Orlin, James B.},
  year = {1993},
  publisher = {Prentice Hall},
  address = {Englewood Cliffs, NJ},
  isbn = {978-0-13-617549-0}
}

@article{Goldstein1998RadarInterferogramFiltering,
  title = {Radar Interferogram Filtering for Geophysical Applications},
  author = {Goldstein, Richard M. and Werner, Charles L.},
  year = {1998},
  journal = {Geophysical Research Letters},
  volume = {25},
  number = {21},
  pages = {4035--4038},
  doi = {10.1029/1998GL900033}
}

@article{Chen2015PersistentScattererInterpolation,
  title = {A Persistent Scatterer Interpolation for Retrieving Accurate Ground
           Deformation over {InSAR}-Decorrelated Agricultural Fields},
  author = {Chen, Jingyi and Zebker, Howard A. and Knight, Rosemary},
  year = {2015},
  journal = {Geophysical Research Letters},
  volume = {42},
  number = {21},
  pages = {9294--9301},
  doi = {10.1002/2015GL065031}
}

@article{Wang2022AccuratePersistentScatterer,
  title = {Accurate Persistent Scatterer Identification Based on Phase
           Similarity of Radar Pixels},
  author = {Wang, Ke and Chen, Jingyi},
  year = {2022},
  journal = {IEEE Transactions on Geoscience and Remote Sensing},
  volume = {60},
  pages = {1--13},
  doi = {10.1109/TGRS.2022.3210868}
}

@article{Ansari2017SequentialEstimator,
  title = {Sequential Estimator: Toward Efficient {InSAR} Time Series Analysis},
  author = {Ansari, Homa and {De Zan}, Francesco and Bamler, Richard},
  year = {2017},
  journal = {IEEE Transactions on Geoscience and Remote Sensing},
  volume = {55},
  number = {10},
  pages = {5637--5652},
  doi = {10.1109/TGRS.2017.2711037}
}

@article{Pepe2006ExtensionMinimumCostFlow,
  title = {On the Extension of the Minimum Cost Flow Algorithm for Phase
           Unwrapping of Multitemporal Differential {SAR} Interferograms},
  author = {Pepe, Antonio and Lanari, Riccardo},
  year = {2006},
  journal = {IEEE Transactions on Geoscience and Remote Sensing},
  volume = {44},
  number = {9},
  pages = {2374--2383},
  doi = {10.1109/TGRS.2006.873207}
}

@article{Goldstein1988SatelliteRadarInterferometry,
  title = {Satellite Radar Interferometry: Two-Dimensional Phase Unwrapping},
  author = {Goldstein, Richard M. and Zebker, Howard A. and Werner, Charles L.},
  year = {1988},
  journal = {Radio Science},
  volume = {23},
  number = {4},
  pages = {713--720},
  doi = {10.1029/RS023i004p00713}
}

@article{Costantini1998NovelPhaseUnwrapping,
  title = {A Novel Phase Unwrapping Method Based on Network Programming},
  author = {Costantini, Mario},
  year = {1998},
  journal = {IEEE Transactions on Geoscience and Remote Sensing},
  volume = {36},
  number = {3},
  pages = {813--821},
  doi = {10.1109/36.673674}
}

@article{Carballo2000ProbabilisticCostFunctions,
  title = {Probabilistic Cost Functions for Network Flow Phase Unwrapping},
  author = {Carballo, Gustavo F. and Fieguth, Paul W.},
  year = {2000},
  journal = {IEEE Transactions on Geoscience and Remote Sensing},
  volume = {38},
  number = {5},
  pages = {2192--2201},
  doi = {10.1109/36.868877}
}

@inproceedings{Werner2002ProcessingStrategiesPhase,
  title = {Processing strategies for phase unwrapping for INSAR applications},
  author = {Werner, C and Wegm{\"u}ller, Urs and Strozzi, Tazio and Wiesmann, A},
  booktitle = {proceedings of the European conference on synthetic aperture radar (EUSAR 2002)},
  volume = {1},
  pages = {353--356},
  year = {2002}
}

@article{Itoh1982AnalysisPhaseUnwrapping,
  title = {Analysis of the Phase Unwrapping Algorithm},
  author = {Itoh, Kazuyoshi},
  year = 1982,
  month = jul,
  journal = {Applied Optics},
  volume = {21},
  number = {14},
  pages = {2470},
  issn = {0003-6935, 1539-4522},
  doi = {10.1364/AO.21.002470},
  urldate = {2026-07-27},
  copyright = {https://doi.org/10.1364/OA\_License\_v1\#VOR},
  langid = {english}
}

@article{Chen2002PhaseUnwrappingLarge,
  title = {Phase Unwrapping for Large {SAR} Interferograms: Statistical
           Segmentation and Generalized Network Models},
  author = {Chen, Curtis W. and Zebker, Howard A.},
  year = {2002},
  journal = {IEEE Transactions on Geoscience and Remote Sensing},
  volume = {40},
  number = {8},
  pages = {1709--1719},
  doi = {10.1109/TGRS.2002.802453}
}

@article{Lee1994IntensityPhaseStatistics,
  title = {Intensity and Phase Statistics of Multilook Polarimetric and
           Interferometric {SAR} Imagery},
  author = {Lee, Jong-Sen and Hoppel, Karl W. and Mango, Stephen A. and
            Miller, Allen R.},
  year = {1994},
  journal = {IEEE Transactions on Geoscience and Remote Sensing},
  volume = {32},
  number = {5},
  pages = {1017--1028},
  doi = {10.1109/36.312890}
}

@article{Yu2019PhaseUnwrappingReview,
  title = {Phase Unwrapping in {InSAR}: A Review},
  author = {Yu, Hanwen and Lan, Yang and Yuan, Zhihui and Xu, Jianyu and
            Lee, Hyongki},
  year = {2019},
  journal = {IEEE Geoscience and Remote Sensing Magazine},
  volume = {7},
  number = {1},
  pages = {40--58},
  doi = {10.1109/MGRS.2018.2873644}
}

@article{Pepin2024AliasingInSAR,
  title = {Aliasing in {InSAR} 2-{D} Phase Unwrapping and Time Series},
  author = {Pepin, Karissa and Zebker, Howard},
  year = {2024},
  journal = {IEEE Transactions on Geoscience and Remote Sensing},
  volume = {62},
  pages = {1--17},
  doi = {10.1109/TGRS.2024.3359482}
}

@techreport{Wu2017ParallelASSP,
  title = {Two-Dimensional Phase Unwrapping with Parallel Augmenting
           Successive Shortest Paths},
  author = {Wu, Xiaoqing},
  institution = {Jet Propulsion Laboratory, California Institute of Technology},
  type = {Interoffice Memorandum},
  number = {IOM 334-XWU},
  year = {2017},
  month = may,
  note = {Internal memorandum; not publicly published}
}

@article{Dial1969Algorithm360,
  title = {Algorithm 360: Shortest-Path Forest with Topological Ordering},
  author = {Dial, Robert B.},
  year = {1969},
  journal = {Communications of the ACM},
  volume = {12},
  number = {11},
  pages = {632--633},
  doi = {10.1145/363269.363610}
}

@misc{isce3,
  title = {{ISCE3}: {InSAR} Scientific Computing Environment 3},
  howpublished = {\url{https://github.com/isce-framework/isce3}},
  note = {Reference implementations of {PHASS} and {ICU} used in the
          benchmark.}
}

@misc{nisar,
  title = {{NISAR}: {NASA-ISRO} Synthetic Aperture Radar Mission},
  author = {{NASA Jet Propulsion Laboratory}},
  howpublished = {\url{https://nisar.jpl.nasa.gov}},
  year = {2026}
}

@misc{nisar_l2_gunw_beta_v1,
  title = {{NISAR Level 2 Geocoded Unwrapped Interferogram (GUNW) Beta Version 1}},
  author = {{NASA Jet Propulsion Laboratory}},
  howpublished = {Alaska Satellite Facility Distributed Active Archive Center,
                  \url{https://nisar-docs.asf.alaska.edu/aws-s3-access/}},
  year = {2026},
  note = {Collection short name: NISAR\_L2\_GUNW\_BETA\_V1}
}

@misc{nisar_l2_gunw_provisional_v1,
  title = {{NISAR Level 2 Geocoded Unwrapped Interferogram (GUNW) Provisional Version 1}},
  author = {{NASA Jet Propulsion Laboratory}},
  howpublished = {Alaska Satellite Facility Distributed Active Archive Center,
                  \url{https://nisar-docs.asf.alaska.edu/aws-s3-access/}},
  year = {2026},
  note = {Collection short name: NISAR\_L2\_GUNW\_PROVISIONAL\_V1}
}

@article{Staniewicz2024DolphinPythonPackage,
  title = {Dolphin: {{A Python}} Package for Large-Scale {{InSAR PS}}/{{DS}} Processing},
  shorttitle = {Dolphin},
  author = {Staniewicz, Scott J. and Mirzaee, Sara and Gunter, Geoffrey M. and {Oliver-Cabrera}, Talib and Havazli, Emre and Fattahi, Heresh},
  year = 2024,
  month = nov,
  journal = {Journal of Open Source Software},
  volume = {9},
  number = {103},
  pages = {6997},
  issn = {2475-9066},
  doi = {10.21105/joss.06997},
  urldate = {2025-04-24},
  langid = {english}
}

@article{Staniewicz2026NearRealTimeInSARPhase,
  title = {Near-{{Real-Time InSAR Phase Estimation}} for {{Large-Scale Surface Displacement Monitoring}}},
  author = {Staniewicz, Scott and Mirzaee, Sara and Fattahi, Heresh and {Oliver-Cabrera}, Talib and Havazli, Emre and Gunter, Geoffrey and Jeon, Se-Yeon and Bato, Mary Grace and Kim, Jinwoo and Sangha, Simran S. and Chapman, Bruce D. and Handwerger, Alexander L. and Govorcin, Marin and Agram, Piyush and Bekaert, David P. S.},
  year = 2026,
  journal = {IEEE Transactions on Geoscience and Remote Sensing},
  volume = {64},
  note = {Art. no. 5207116},
  issn = {1558-0644},
  doi = {10.1109/TGRS.2026.3685947},
  urldate = {2026-07-08}
}

@misc{spurt,
  title = {spurt: Spatial and temporal phase unwrapping for {InSAR} time series},
  author = {{ISCE framework contributors}},
  howpublished = {\url{https://github.com/isce-framework/spurt}},
  year = {2024}
}

@misc{operadisp,
  title = {{OPERA} {North America} Surface Displacement Product Suite},
  author = {{OPERA Project, Jet Propulsion Laboratory}},
  howpublished = {\url{https://www.jpl.nasa.gov/go/opera}},
  year = {2025}
}

@article{Thollard2021FLATSIMForMTerLArgeScale,
  title = {{{FLATSIM}}: {{The ForM}}@{{Ter LArge-Scale Multi-Temporal Sentinel-1 InterferoMetry Service}}},
  shorttitle = {{{FLATSIM}}},
  author = {Thollard, Franck and Clesse, Dominique and Doin, Marie-Pierre and Donadieu, Jo{\"e}lle and Durand, Philippe and Grandin, Rapha{\"e}l and Lasserre, C{\'e}cile and Laurent, Christophe and {Deschamps-Ostanciaux}, Emilie and Pathier, Erwan and Pointal, Elisabeth and Proy, Catherine and Specht, Bernard},
  year = 2021,
  month = sep,
  journal = {Remote Sensing},
  volume = {13},
  number = {18},
  pages = {3734},
  issn = {2072-4292},
  doi = {10.3390/rs13183734},
  urldate = {2025-09-15},
  copyright = {https://creativecommons.org/licenses/by/4.0/},
  langid = {english}
}

@article{Torres2012GMESSentinel1Mission,
  title = {{{GMES Sentinel-1}} Mission},
  author = {Torres, Ramon and Snoeij, Paul and Geudtner, Dirk and Bibby, David and Davidson, Malcolm and Attema, Evert and Potin, Pierre and Rommen, Bj{\"O}rn and Floury, Nicolas and Brown, Mike and Traver, Ignacio Navas and Deghaye, Patrick and Duesmann, Berthyl and Rosich, Betlem and Miranda, Nuno and Bruno, Claudio and L'Abbate, Michelangelo and Croci, Renato and Pietropaolo, Andrea and Huchler, Markus and Rostan, Friedhelm},
  year = 2012,
  month = may,
  journal = {Remote Sensing of Environment},
  series = {The {{Sentinel Missions}} - {{New Opportunities}} for {{Science}}},
  volume = {120},
  pages = {9--24},
  issn = {0034-4257},
  doi = {10.1016/j.rse.2011.05.028},
  urldate = {2026-07-27}
}

@inproceedings{Geudtner2021CopernicusSentinel1Next,
  title = {Copernicus {{Sentinel-1 Next Generation Mission}}},
  booktitle = {2021 {{IEEE International Geoscience}} and {{Remote Sensing Symposium IGARSS}}},
  author = {Geudtner, Dirk and Tossaint, Michel and Davidson, Malcolm and Torres, Ramon},
  year = 2021,
  month = jul,
  pages = {874--876},
  publisher = {IEEE},
  address = {Brussels, Belgium},
  doi = {10.1109/IGARSS47720.2021.9554226},
  urldate = {2026-08-06},
  copyright = {https://ieeexplore.ieee.org/Xplorehelp/downloads/license-information/IEEE.html},
  isbn = {978-1-6654-0369-6},
  langid = {english}
}

@article{Yunjun2019MintPy,
  author = {Yunjun, Zhang and Fattahi, Heresh and Amelung, Falk},
  title = {Small baseline {InSAR} time series analysis: Unwrapping error correction and noise reduction},
  journal = {Computers \& Geosciences},
  volume = {133},
  pages = {104331},
  year = {2019},
  doi = {10.1016/j.cageo.2019.104331}
}

@article{Ross2019Ridgecrest,
  author = {Ross, Zachary E. and Idini, Benjam{\'i}n and Jia, Zhe and Stephenson, Oliver L. and Zhong, Minyan and Wang, Xin and Zhan, Zhongwen and Simons, Mark and Fielding, Eric J. and Yun, Sang-Ho and Hauksson, Egill and Moore, Angelyn W. and Liu, Zhen and Jung, Jungkyo},
  title = {Hierarchical interlocked orthogonal faulting in the 2019 {Ridgecrest} earthquake sequence},
  journal = {Science},
  volume = {366},
  number = {6463},
  pages = {346--351},
  year = {2019},
  doi = {10.1126/science.aaz0109}
}

@article{Calabro2010PortugueseBend,
  title = {An Examination of Seasonal Deformation at the {Portuguese Bend}
           Landslide, Southern {California}, Using Radar Interferometry},
  author = {Calabro, M. D. and Schmidt, D. A. and Roering, J. J.},
  year = {2010},
  journal = {Journal of Geophysical Research: Earth Surface},
  volume = {115},
  number = {F2},
  pages = {F02020},
  doi = {10.1029/2009JF001314}
}

@article{Xu2020CoseismicDisplacementsSurface,
  title = {Coseismic {{Displacements}} and {{Surface Fractures}} from {{Sentinel}}-1 {{InSAR}}: 2019 {{Ridgecrest Earthquakes}}},
  shorttitle = {Coseismic {{Displacements}} and {{Surface Fractures}} from {{Sentinel}}-1 {{InSAR}}},
  author = {Xu, Xiaohua and Sandwell, David T. and Smith-Konter, Bridget},
  year = {2020},
  month = jan,
  journal = {Seismological Research Letters},
  volume = {91},
  number = {4},
  pages = {1979--1985},
  issn = {0895-0695},
  doi = {10.1785/0220190275},
  urldate = {2021-08-12}
}

@article{Feigl2009MethodModellingRadar,
  title = {A Method for Modelling Radar Interferograms without Phase Unwrapping: Application to the {{M}} 5 {{Fawnskin}}, {{California}} Earthquake of 1992 {{December}} 4},
  author = {Feigl, Kurt L. and Thurber, Clifford H.},
  year = {2009},
  journal = {Geophysical Journal International},
  volume = {176},
  number = {2},
  pages = {491--504},
  doi = {10.1111/j.1365-246X.2008.03881.x}
}

@article{Lazecky2020LiCSARAutomaticInSAR,
  title = {{{LiCSAR}}: {{An Automatic InSAR Tool}} for {{Measuring}} and {{Monitoring Tectonic}} and {{Volcanic Activity}}},
  shorttitle = {{{LiCSAR}}},
  author = {Lazeck{\'y}, Milan and Spaans, Karsten and Gonz{\'a}lez, Pablo J. and Maghsoudi, Yasser and Morishita, Yu and Albino, Fabien and Elliott, John and Greenall, Nicholas and Hatton, Emma and Hooper, Andrew and Juncu, Daniel and McDougall, Alistair and Walters, Richard J. and Watson, C. Scott and Weiss, Jonathan R. and Wright, Tim J.},
  year = 2020,
  month = jan,
  journal = {Remote Sensing},
  volume = {12},
  number = {15},
  pages = {2430},
  publisher = {Multidisciplinary Digital Publishing Institute},
  issn = {2072-4292},
  doi = {10.3390/rs12152430},
  urldate = {2022-03-09},
  copyright = {http://creativecommons.org/licenses/by/3.0/},
  langid = {english}
}

\clearpage
\appendices
\onecolumn
\setcounter{figure}{0}
\setcounter{table}{0}
\renewcommand{\thefigure}{S\arabic{figure}}
\renewcommand{\thetable}{S\arabic{table}}
\renewcommand{\theHfigure}{S\arabic{figure}}
\renewcommand{\theHtable}{S\arabic{table}}

\section{Further Whirlwind--SNAPHU Comparisons from the Provisional NISAR Campaign}\label{sec:supp-lowagree}

Figures~\ref{fig:supp-rank02}--\ref{fig:supp-rank43} show a selection of some of the frames with the largest disagreements between Whirlwind and the production SNAPHU unwrapping among the 6,014 provisional NISAR GUNW products we compared.
Each figure repeats the panel layout of Figure~3 of the main text, showing the input re-wrapped production phase and the GUNW coherence, the output production unwrapped phase, and the Whirlwind unwrapped phase, aligned to a common reference with the production unwrapped phase. The bottom rows show production connected-component labels, Whirlwind connected-component labels, the difference in component coverage, and the integer-cycle difference between the two unwrapped fields.
We note that some of the perceived problems with the production SNAPHU results may be due to misconfiguration of the SNAPHU unwrapper for these specific areas (predominately frames with transmit gaps associated with fixed PRF), rather than problems with the cost or underlying algorithms.
Frames with gaps between subswaths are expected from NISAR's fixed PRF mode, and differences in these frames are due to the implementation of the bridging post-processing.

\begin{figure}[!p]
  \centering
  \includegraphics[width=\textwidth]{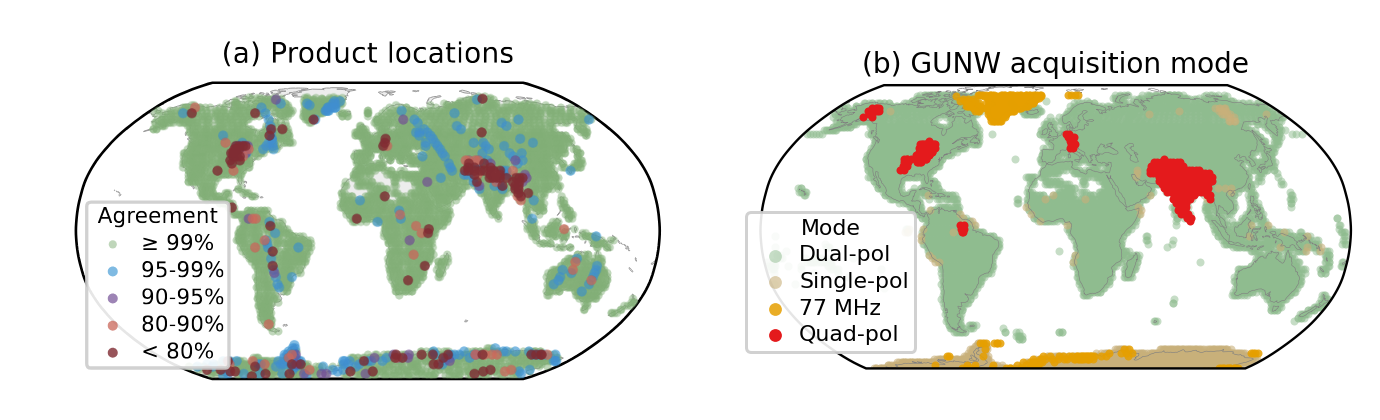}
  \caption{(a) NISAR GUNW products colored by agreement between Whirlwind and the NISAR SNAPHU production unwrapped phase (reproduced from Main Text Figure 5a). 
  (b) Acquisition mode of provisional NISAR GUNW frames (as of September 2026).}
  \label{fig:dissagreement_vs_mode}
\end{figure}

\begin{figure}[!p]
  \centering
  \includegraphics[width=\textwidth]{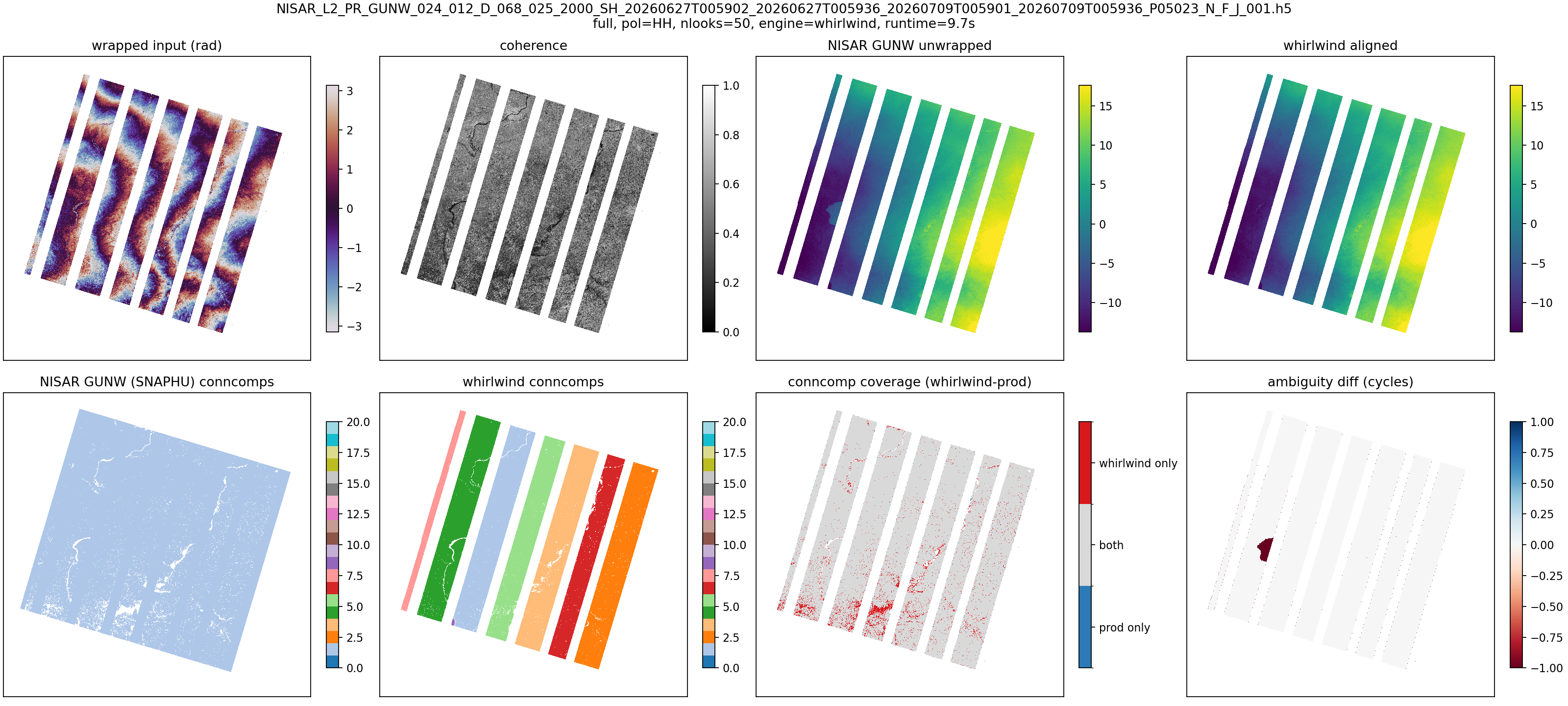}
  \caption{Cycle~24, track~12, frame~68 (descending), a fixed-PRF mode. Note that the nonzero connected component label in the empty subswaths is from a misconfiguration and is fixed in later NISAR data releases.}
  \label{fig:supp-rank02}
\end{figure}

\begin{figure}[!p]
  \centering
  \includegraphics[width=\textwidth]{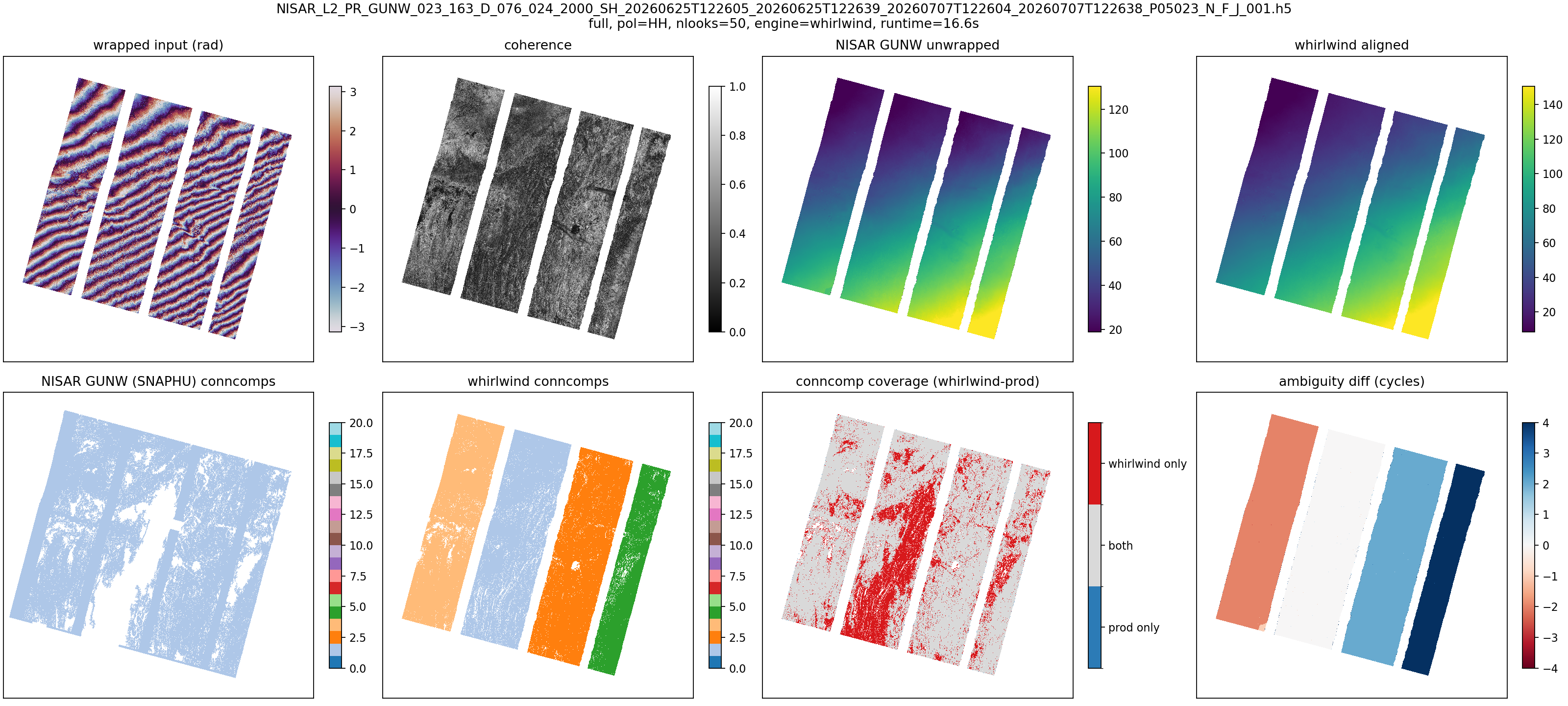}
  \caption{Cycle~23, track~163, frame~76 (descending). The phase ramp, combined with the subswath gaps, leaves a more challenging case for bridging algorithms.}
  \label{fig:supp-rank03}
\end{figure}

\begin{figure}[!p]
  \centering
  \includegraphics[width=\textwidth]{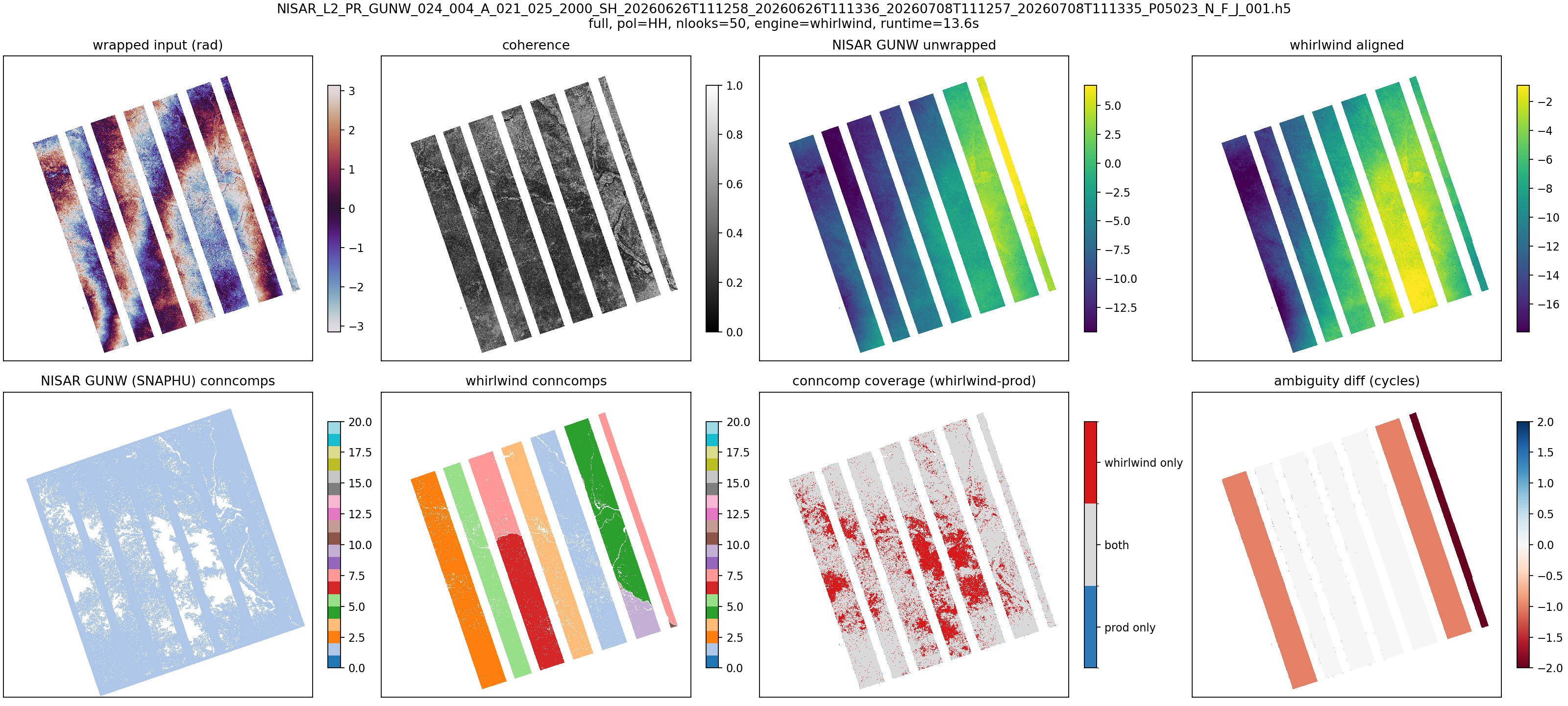}
    \caption{Cycle~24, track~4, frame~21 (ascending).}
  \label{fig:supp-rank06}
\end{figure}

\begin{figure}[!p]
  \centering
  \includegraphics[width=\textwidth]{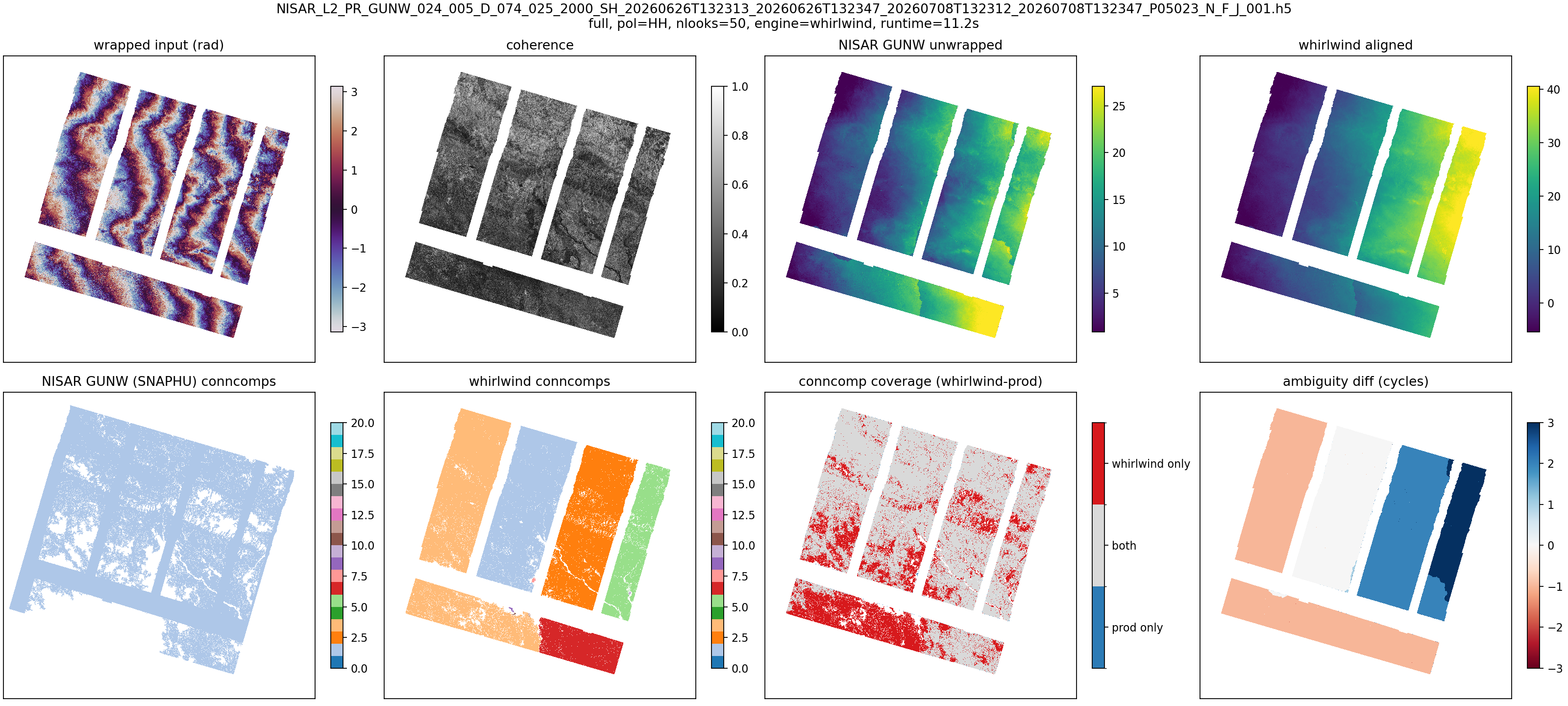}
    \caption{Cycle~24, track~5, frame~74 (descending). }
  \label{fig:supp-rank07}
\end{figure}

\begin{figure}[!p]
  \centering
  \includegraphics[width=\textwidth]{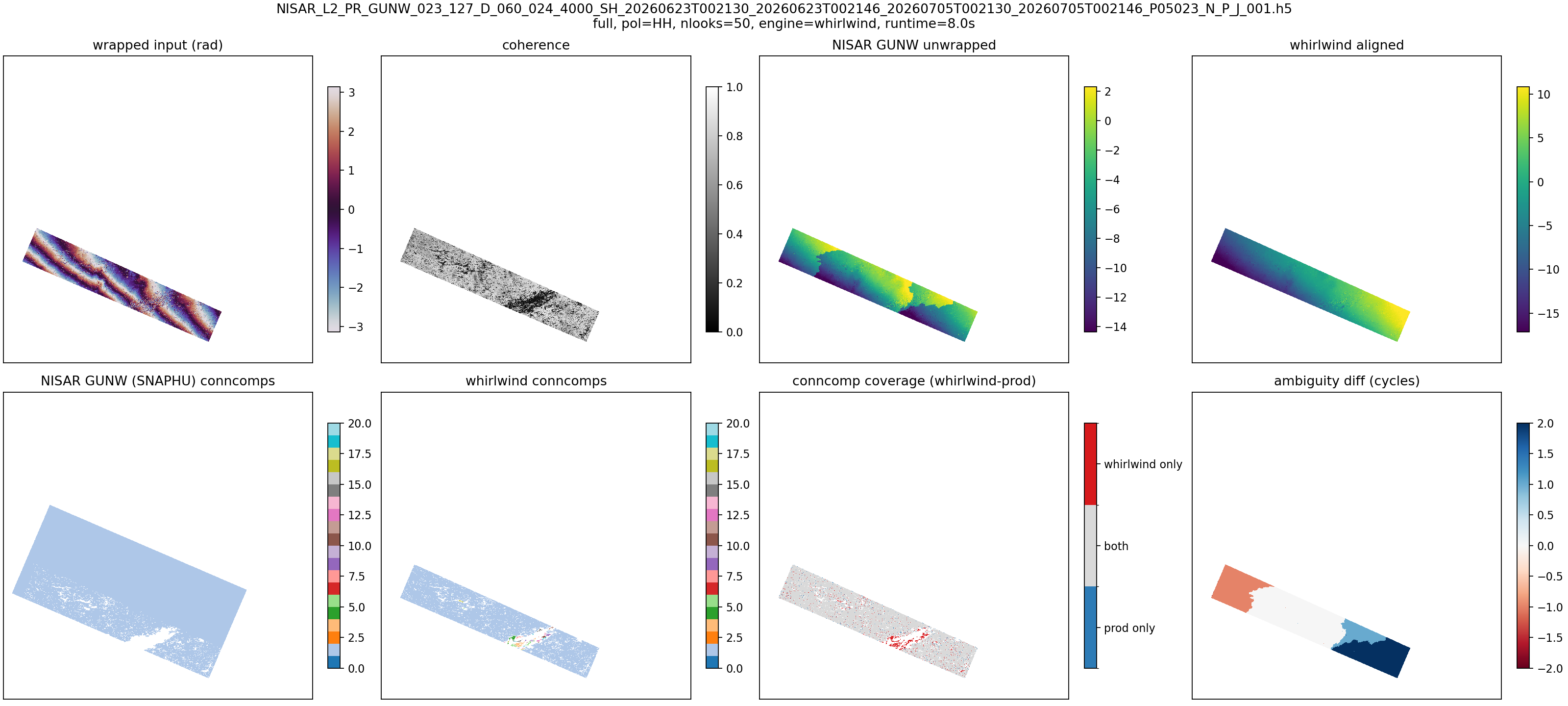}
    \caption{Cycle~23, track~127, frame~60 (descending). }
  \label{fig:supp-rank10}
\end{figure}

\begin{figure}[!p]
  \centering
  \includegraphics[width=\textwidth]{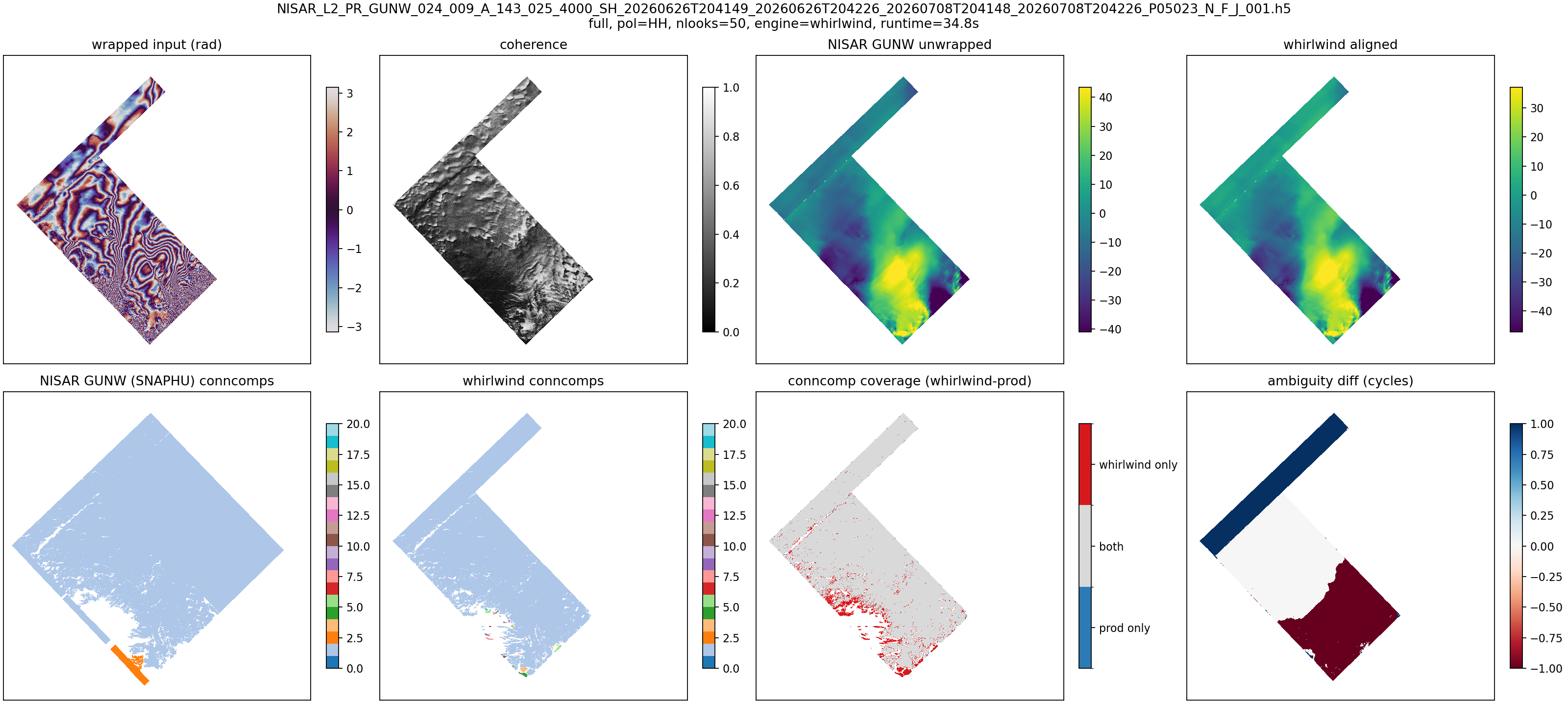}
    \caption{Cycle~24, track~9, frame~143 (ascending), Antarctica. }
  \label{fig:supp-rank13}
\end{figure}

\begin{figure}[!p]
  \centering
  \includegraphics[width=\textwidth]{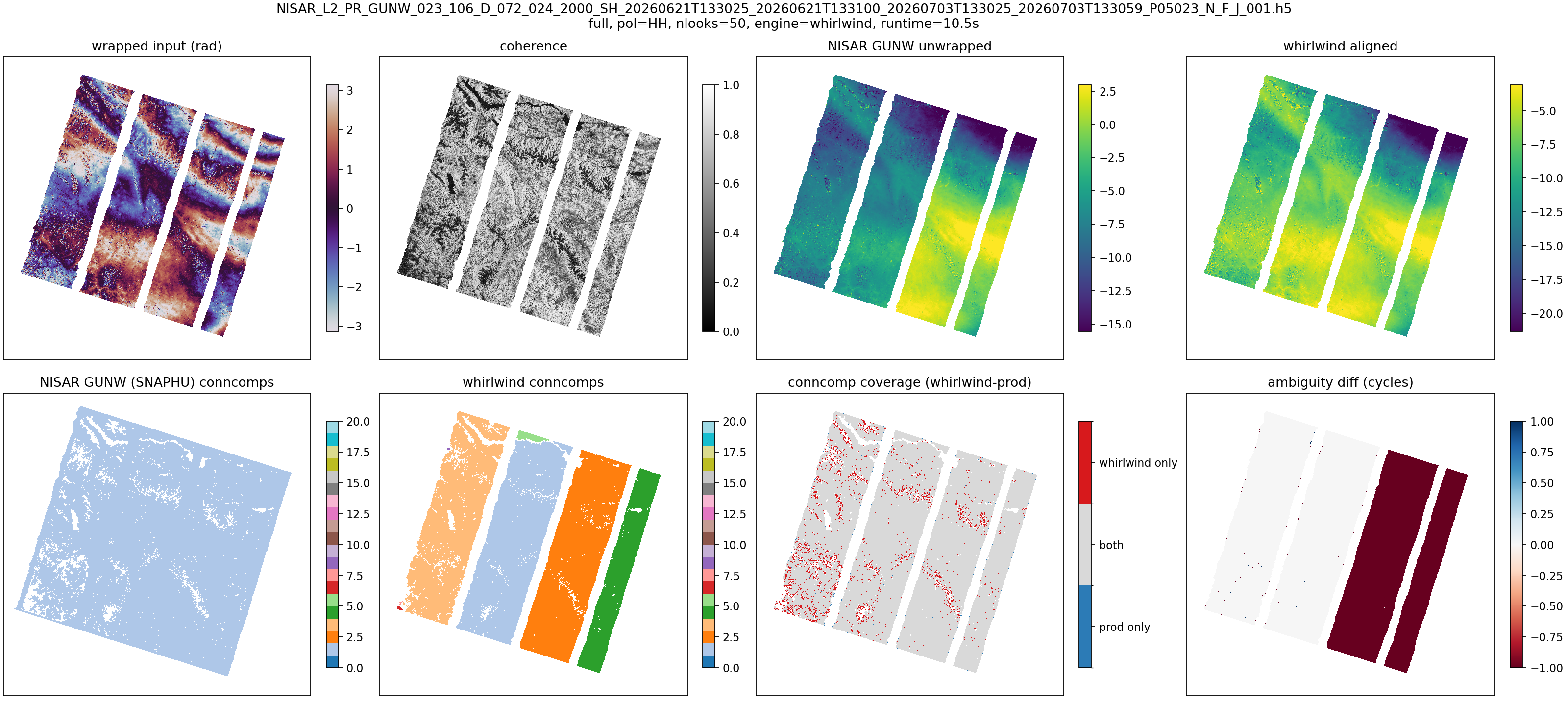}
    \caption{Cycle~23, track~106, frame~72 (descending). }
  \label{fig:supp-rank30}
\end{figure}

\begin{figure}[!p]
  \centering
  \includegraphics[width=\textwidth]{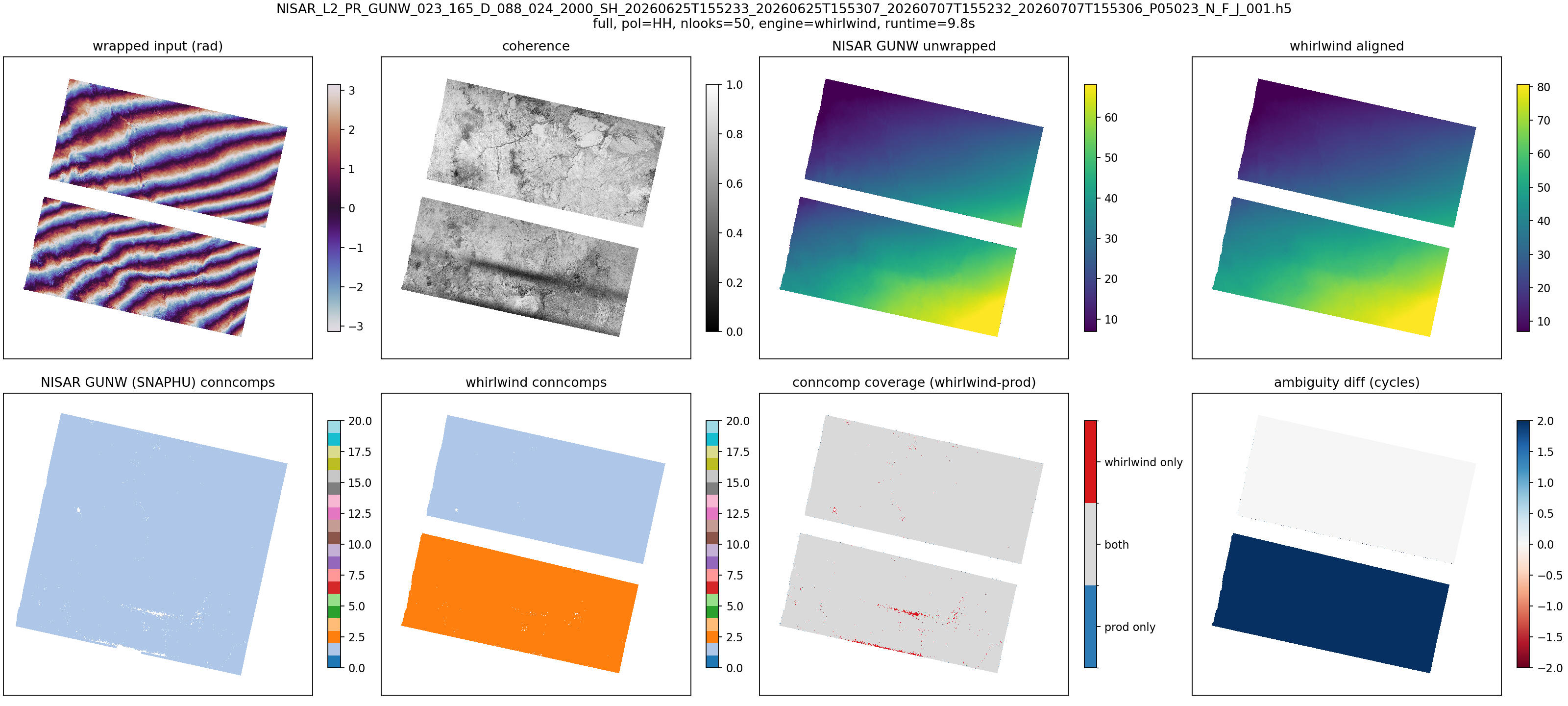}
    \caption{Cycle~23, track~165, frame~88 (descending). }
  \label{fig:supp-rank31}
\end{figure}

\begin{figure}[!p]
  \centering
  \includegraphics[width=\textwidth]{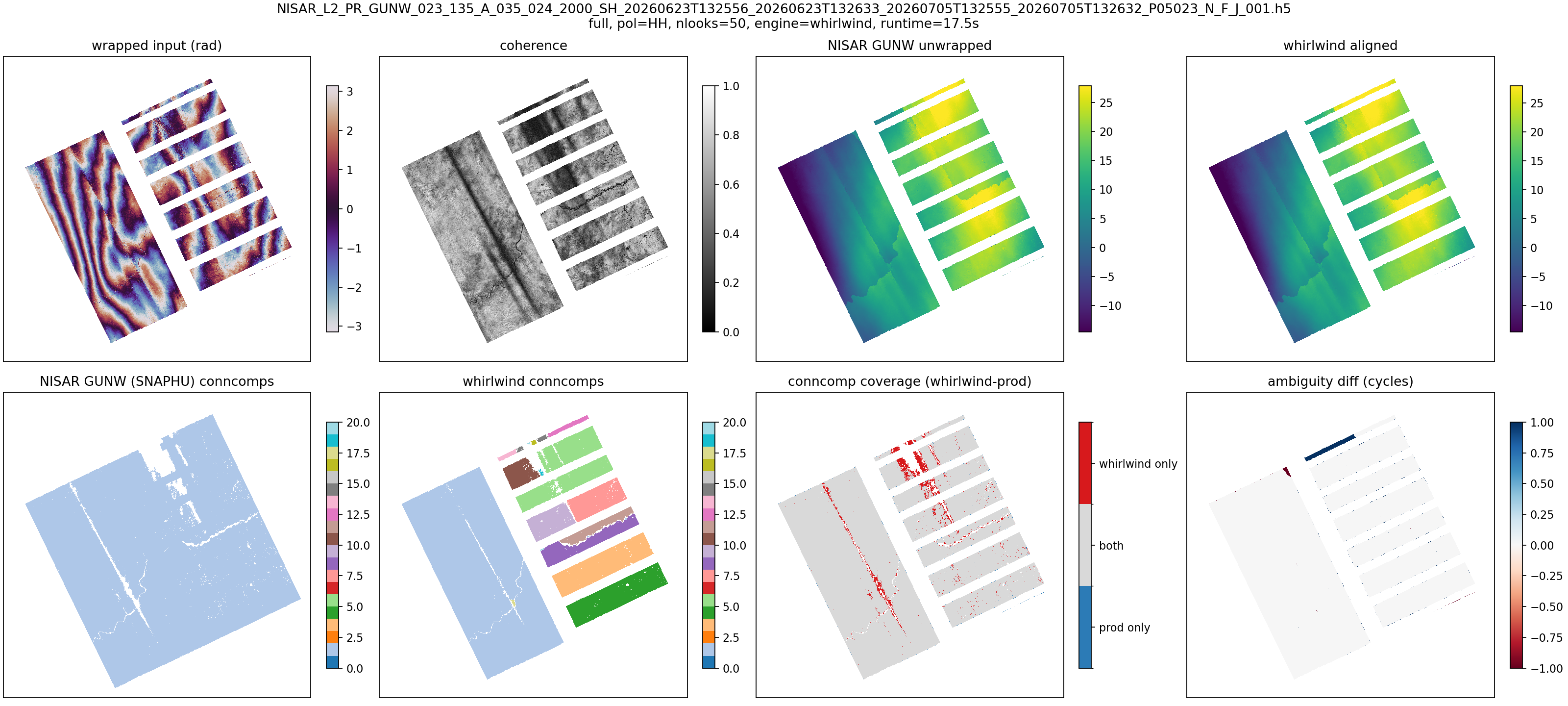}
    \caption{Cycle~23, track~135, frame~35 (ascending). }
  \label{fig:supp-rank33}
\end{figure}

\begin{figure}[!p]
  \centering
  \includegraphics[width=\textwidth]{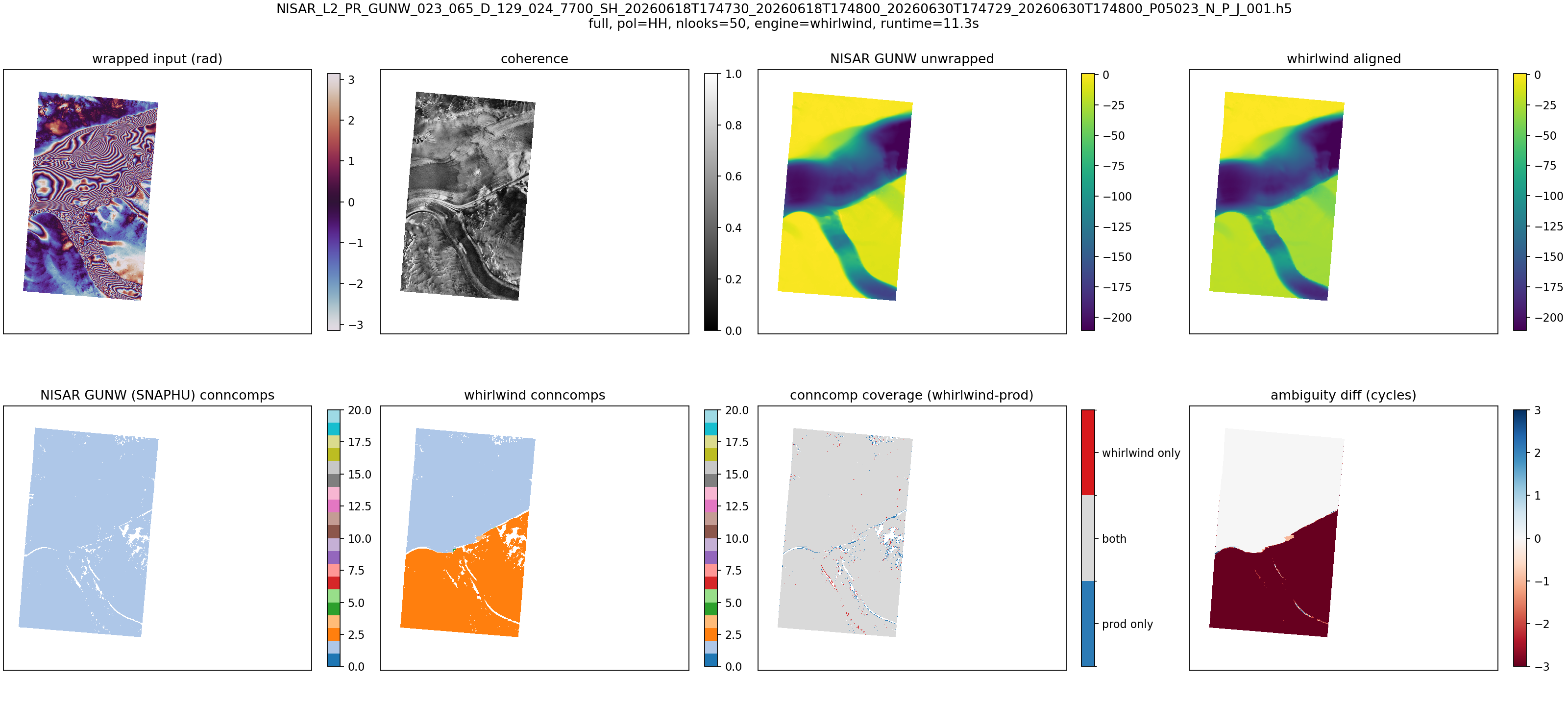}
    \caption{Cycle~23, track~65, frame~129 (descending), Antarctica. }
  \label{fig:supp-rank38}
\end{figure}

\begin{figure}[!p]
  \centering
  \includegraphics[width=\textwidth]{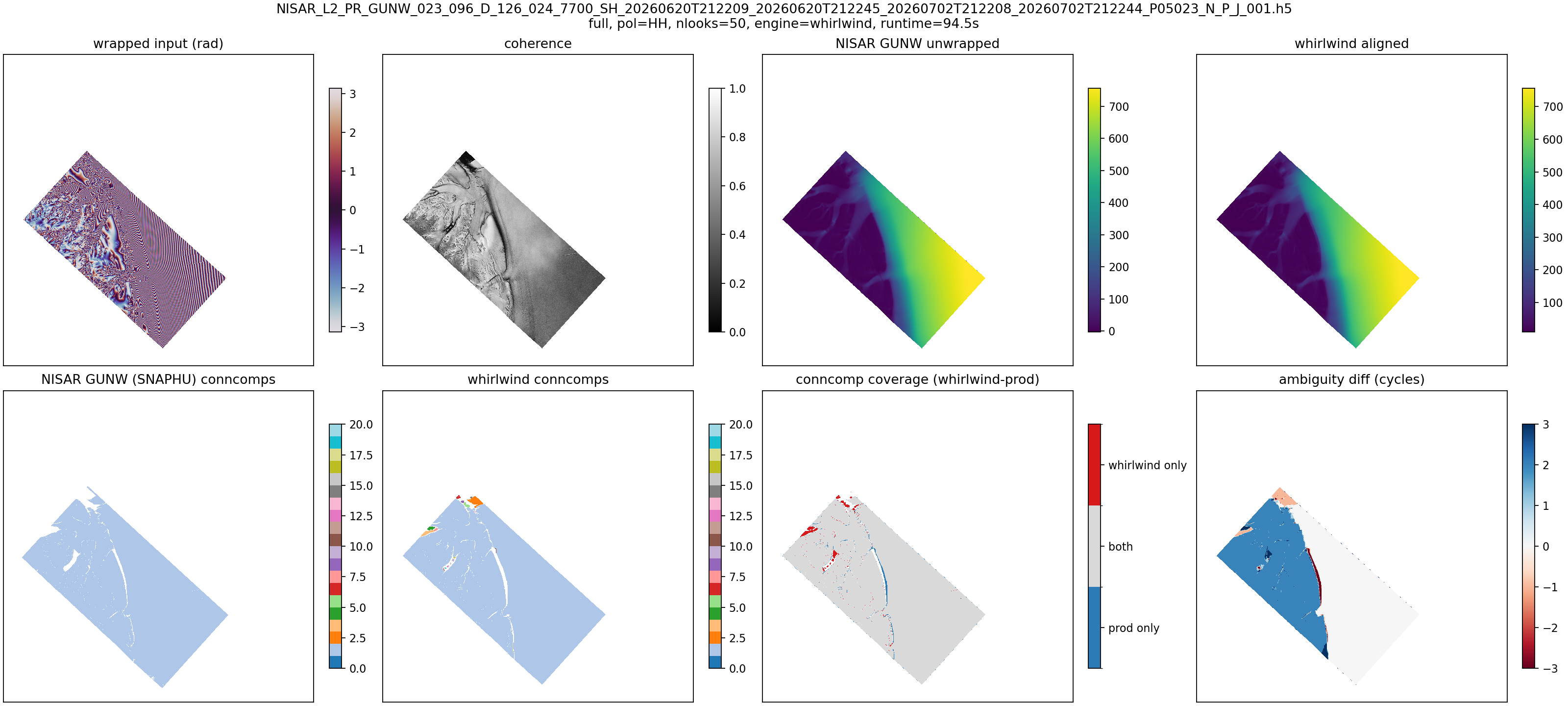}
    \caption{Cycle~23, track~96, frame~126 (descending), Antarctica. }
  \label{fig:supp-rank43}
\end{figure}

\end{document}